\documentclass[submission, Phys]{SciPost}
\usepackage[T1]{fontenc}        
\usepackage[utf8]{inputenc}
\usepackage[english]{babel}

\usepackage[protrusion,
            expansion
            ]{microtype} 

\usepackage{booktabs}    

\usepackage[font=small,labelfont=bf]{caption} 
\usepackage{subcaption}

\usepackage{enumitem}

\usepackage{authblk}    
\usepackage{appendix}
\usepackage{fancyhdr}

\usepackage{braket}
\usepackage{xspace}
\usepackage{mathdots}
\usepackage{stackrel}
\usepackage{xcolor}
\usepackage{mathtools}
\usepackage{graphicx}
\usepackage{amsmath,          
            amssymb,          
            amsthm}          
\usepackage{xfrac}
\usepackage{comment}

\usepackage{mathrsfs}
\usepackage{bbm}      
\usepackage{tikz}

\definecolor{mycolor1}{HTML}{1F7A8C}
\definecolor{mycolor2}{HTML}{2D3362}
\definecolor{mycolor3}{HTML}{BF1363}
\definecolor{mycolor4}{HTML}{F39237}
\definecolor{mycolor5}{HTML}{2D8B6A}

\usepackage{cleveref}
\crefname{figure}{Fig.}{Figs.}
\crefname{equation}{Eq.}{Eqs.}
\crefname{section}{Sec.}{Secs.}

\allowdisplaybreaks

\theoremstyle{definition}

\theoremstyle{remark}

\newcommand{\ii}{\mathrm{i}}

\newcommand{\dd}{\mathrm{d}}

\newcommand{\px}{\sigma^{x}}
\newcommand{\py}{\sigma^{y}}
\newcommand{\pz}{\sigma^{z}}

\newcommand{\abs}[1]{\left\vert#1\right\vert}

\newcommand{\lr}[1]{\left( {#1} \right)} 
\newcommand{\slr}[1]{\left[{#1} \right]}
\newcommand{\clr}[1]{\left\{{#1} \right\}}

\DeclareMathOperator{\sgn}{sgn}
\DeclareMathOperator{\Tr}{Tr}   
\DeclareMathOperator{\Li}{Li}   

\DeclareMathOperator{\imag}{Im}

\title{\textbf{Universality of equilibration dynamics after quantum quenches}}

\author[1,2]{Vincenzo Alba}
\author[1,2]{Sanam Azarnia}
\author[3]{Gianluca Lagnese}
\author[1,2]{Federico Rottoli}

\affil[1]{\textit{Dipartimento di Fisica dell’Universit\`a di Pisa, Largo B. Pontecorvo 3, I-56127 Pisa, Italy.}}
\affil[2]{\textit{INFN Sezione di Pisa, Largo B. Pontecorvo 3, I-56127 Pisa, Italy.}}

\affil[3]{\textit{Jo\v{z}ef Stefan Institute, Jamova cesta 39, 1000 Ljubljana, Slovenia.}}

\date{}

\begin{document}
\maketitle

\begin{abstract}
We investigate the distribution of the eigenvalues of the reduced density matrix (entanglement spectrum) after a global quantum quench. We show that in an appropriate scaling limit the lower part of the entanglement spectrum exhibits ``universality''. In the scaling limit and at asymptotically long times  the distribution of the entanglement spectrum depends on two parameters that can be determined from the R\'enyi entropies. We show that two typical scenarios occur.  
In the first one, the distribution of the entanglement spectrum levels is similar to the one describing the ground-state entanglement spectrum in Conformal Field Theories.  In the second scenario, the lower levels of the entanglement spectrum are highly degenerate and their distribution is given by a series of Dirac deltas. We benchmark our analytical results in free-fermion chains, such as the transverse field Ising chain and the XX chain, in the rule $54$ chain, and in Bethe ansatz solvable spin models.  
\end{abstract}

\tableofcontents

\section{Introduction}\label{sec:intro}

\begin{figure}
    \centering
          \includegraphics[width=.7\linewidth]{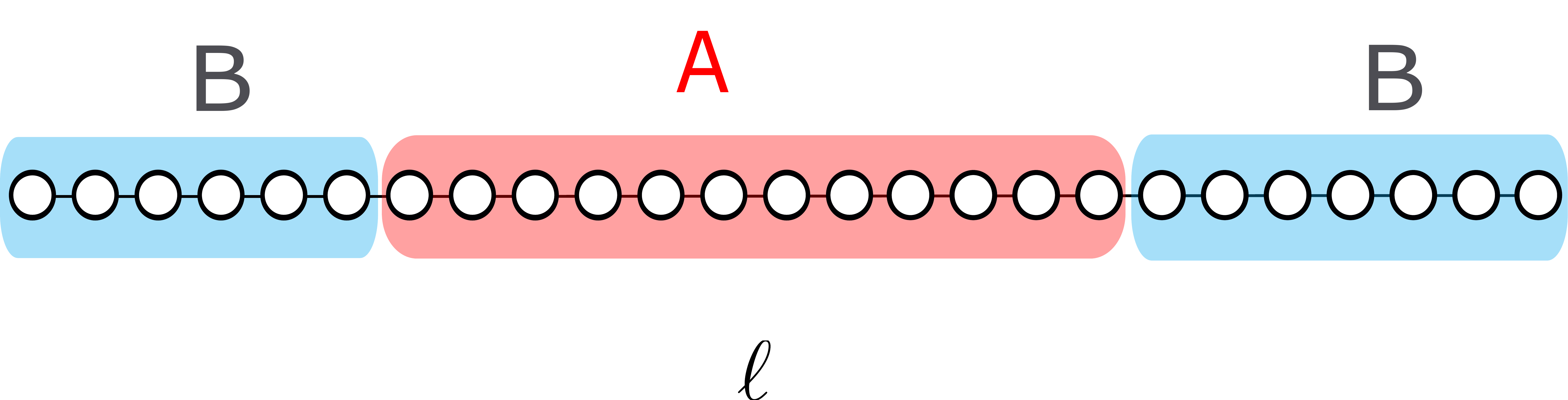}
    \caption{Cartoon of the setup that employed in this work. A subsystem $A$ of length $\ell$ is embedded in an infinite chain. The whole system undergoes unitary dynamics with an Hamiltonian $H$. We are interested in structure of the eigenvalues of the reduced density matrix $\rho_A$ (entanglement spectrum) in the hydrodynamic limit $t,\ell\to\infty$. We consider global quenches, i.e., quench protocols giving rise to linear entanglement growth at short times $t/\ell\ll1$ (short-time regime)  and volume-law entanglement in the long-time regime at $t/\ell\gg1$.  
    }
    \label{fig:cartoon}
\end{figure}

Understanding the onset of equilibration in quantum many-body systems is of paramount importance
in quantum statistical mechanics. 
Isolated quantum many-body systems undergoing unitary dynamics after a quantum quench~\cite{calabrese2016introduction} remain globally in a pure state. Yet, any small subregion at long times typically equilibrates, and it is described by a finite-entropy statistical ensemble.
Hence, the reduced density matrix $\rho_A$ of a subregion $A$ of size $\ell$ (see Fig.~\ref{fig:cartoon}) and its spectrum are crucial to understand equilibration.  
From $\rho_A$ one can define the so-called R\'enyi entropies $S_\alpha$ as 
\begin{equation}
\label{eq:renyi-intro}
S_\alpha=\frac{1}{1-\alpha}\ln\mathrm{Tr}(\rho_A^\alpha). 
\end{equation}
In the limit $\alpha\to1$ one obtains the von Neumann entanglement entropy $S=-\mathrm{Tr}\rho_A\ln(\rho_A)$. 
Both the R\'enyi and the von Neumann entanglement entropies are proper measures of the entanglement between $A$ and its complement. 
In the long-time regime  defined as $t\gg \ell$, with $\ell$ large, all the R\'enyi entropies exhibit a volume-law scaling.  The density of von Neumann entanglement entropy  becomes the same as the density of the thermodynamic entropy of the statistical ensemble describing the steady state. In the short time regime defined as $t\to\infty$ with $t\ll \ell$, $S_\alpha$ increase linearly with time. In integrable systems the identification between entanglement and thermodynamic entropy is at the heart of the so-called quasiparticle picture for entanglement spreading, which allows one to describe the full-time dynamics of the von Neumann entropy~\cite{calabrese-2005,fagotti2008evolution,alba2017entanglement}. Within the quasiparticle picture the entanglement dynamics is understood in terms  of the ballistic propagation of \emph{entangled} pairs of quasiparticles (see Refs.~\cite{alba2018entanglementand,bertini2018entanglement,alba2019entanglement} for the application to dynamics from inhomogeneous states). However, while for integrable free-fermion and free-boson systems the quasiparticle picture applies to both the von Neumann and the R\'enyi entropies~\cite{fagotti2008evolution,alba2018entanglement}, it fails to describe the growth of R\'enyi entropies in interacting integrable ones.  Still, a hydrodynamic formula is available for the slope of their linear growth at short times~\cite{klobas2021exact,klobas2022growth}  and the saturation value at long times~\cite{alba2017quench,alba2017renyi,alba2019towards,mestyan2018renyi}. In principle, from the knowledge of the R\'enyi entropies one can reconstruct the full distribution of the eigenvalues $\lambda_i$ of $\rho_A$ (with $-\ln(\lambda_i)$ forming the so-called entanglement spectrum~\cite{li2008entanglement}), as it was done in Ref.~\cite{calabrese2008entanglement} for the ground-state entanglement spectrum of systems described by a Conformal Field Theory (CFT) (see also Ref.~\cite{okunishi1999universal} and~\cite{alba2017entanglementspectrum} for the case of gapped Hamiltonians). A similar strategy was employed in Ref.~\cite{ruggiero2016negativity} (see also~\cite{shapourian2019twisted} and~\cite{mbeng2017negativity}) to obtain the distribution of the eigenvalues of the partially transposed reduced density matrix. While it is clear that the distribution of the entanglement spectrum contains much more information on equilibration and thermalization than the R\'enyi entropies, its study is challenging. Although it is possible to investigate quantum quenches in CFTs obtaining the dynamics of the entanglement spectrum~\cite{calabrese-2005} analytically, at least under some assumptions on the initial state, the result does not straightforwardly apply to microscopic models because the dynamics involves highly-excited states, which are beyond the CFT description. However, some results are available  for lattice models. For instance,  Ref.~\cite{chang2019evolution} pointed out that the distribution of the entanglement spectrum takes a universal form at early times, and follows a  Marchenko-Pastur distribution at long times (as the entanglement spectrum of excited states~\cite{yang2015two}). Moreover, Ref.~\cite{surace2020operator} studied the dynamics of the gaps in the levels of the entanglement spectrum after quenches to the critical transverse field Ising chain (TFIC) showing that the ratios of the gaps contain  information about the underlying CFT (see also Ref.~\cite{robertson2022quenches}).  
The question of universality in out-of-equilibrium entanglement spectra and the applicability of CFT is attracting attention~\cite{kusuki2025universality,wei2025universality}. 
Quite recently, Ref.~\cite{kusuki2025universality} investigated the $\alpha\to0$ limit of the ground-state R\'enyi entropies for $d$-dimensional CFTs, which allows to obtain the distribution of the large eigenvalues of the modular Hamiltonian. 

Here we characterize the  distribution of the lower-part of the entanglement spectrum after global quantum quenches in one-dimensional systems, although our results are likely to apply also in higher-dimensional systems. We consider both the short-time and the long-time regimes, as defined above. Our main result is that in both regimes the lower-part of the entanglement spectrum exhibits universality at long times, and it can be characterized analytically. 
Precisely, the distribution $P(\lambda)$ of the levels exhibits the universal behavior as 
\begin{equation}
\label{eq:P-intro}
P(\lambda)=\delta(\lambda-\lambda_\mathrm{m})+b r_1^{1/2}\frac{I_1(2 r_1\xi)}{\lambda\xi},\quad\mathrm{with}\,\,\xi=\sqrt{b\ln(\lambda_\mathrm{m}/\lambda)},
\end{equation}
%
%
%
where $0\le \lambda_\mathrm{m}\le1$ is the largest eigenvalue of the reduced density matrix, $b=-\ln(\lambda_\mathrm{m})$ is the so-called single-copy entanglement, and $r_1$ is determined from the behavior of the R\'enyi entropies $S_\alpha$ in the large $\alpha$ limit.  Precisely, we assume that $-\ln(M_\alpha)/\mathcal{L}=-\ln(\mathrm{Tr}\rho_A^\alpha)/\mathcal{L}=a_0\alpha+a_1/\alpha+o(1/\alpha)$ at large $\alpha$, with $a_1\ne 0$. Here we define $\mathcal{L}=\ell,2t$ in the long-time and short-time regime, respectively. 
In \cref{eq:P-intro} $I_1(x)$ is the modified Bessel function of the first kind. \cref{eq:P-intro} holds in both the short-time and long-time regime, provided that  $t,\ell$ are large enough. Notably, \cref{eq:P-intro} is formally the same as that describing the ground-state entanglement spectrum of systems described by a CFT~\cite{calabrese2008entanglement}, apart from the factor $r_1$ (one has $r_1=1$ in any $1+1$ CFT). Indeed, the fact that the system is out-of-equilibrium is fully encoded in the scaling with $\mathcal{L}$ of the largest eigenvalue $\lambda_\mathrm{m}$. Notice that the factor $r_1$ depends on which regime is considered. Moreover, in the short-time and long-time regimes one has $b\simeq t$ and $b\sim \ell$, which means that at fixed $\xi$ upon increasing $t$ or $\ell$, \cref{eq:P-intro} describes eigenvalues $\lambda$ that are closer to $\lambda_\mathrm{m}$. 
It is remarkable that although~\eqref{eq:P-intro} is obtained by truncating the large $\alpha$ expansion of $M_\alpha$, it accurately describes the eigenvalues of $\rho_A$ near $\lambda_\mathrm{m}$. Still, it is important to stress that Eq.~\eqref{eq:P-intro} does not capture the \emph{full} entanglement spectrum, as we will clarify in Section~\ref{sec:theory}. 
It is useful to consider the cumulative distribution function $n(\lambda)$ counting the eigenvalues larger than $\lambda$, and defined as 
\begin{equation}
\label{eq:n-intro}
n(\lambda)=\int_{\lambda}^{\lambda_\mathrm{m}}d\lambda P(\lambda)=I_0(2r_1\xi). 
\end{equation}
In CFT systems one has $r_1=1$ and   \cref{eq:n-intro} is super universal, because it depends only on the central charge of the CFT. In out-of-equilibrium systems, \cref{eq:n-intro} shows that the only non-universal information appearing in $n(\lambda)$ is encoded in $r_1$. It is important to stress that for $a_1=0$, \cref{eq:P-intro} gives  $P(\lambda)=\delta(\lambda-\lambda_{\mathrm{m}})$, which signals that the eigenvalues of $\rho_A$ are degenerate. This scenario corresponds, for instance,  to the situation in which  $M_\alpha/\mathcal{L}=a_0\alpha +\mathcal{O}(e^{-d_1\alpha})$ in the large $\alpha$ limit. By taking into account the exponentially suppressed terms, one can show that $P(\lambda)$ contains a series of Dirac delta functions, and $n(\lambda)$ (cf.~\eqref{eq:n-intro}) exhibits a ``staircase'' structure. For some quenches, the exponentially suppressed terms  are absent, implying that all the eigenvalues in the lower part of the entanglement spectrum are degenerate. To benchmark our results,  we consider several integrable models, such as the transverse field Ising chain (TFIC), the rule $54$ chain~\cite{bobenko1993on}, and the spin-$1/2$ anisotropic Heisenberg chain (XXZ chain), discussing several instances of the different scenarios outlined above. 

The manuscript is organized as follows. In Section~\ref{sec:theory} we derive $P(\lambda)$ starting from the large $\alpha$ expansion of the R\'enyi entropies. In Section~\ref{sec:rule-54} we apply our approach to quenches in the rule $54$ chain. In Section~\ref{sec:ising} we discuss the magnetic field quench in the TFIC. In particular, we show that while quenches with the critical TFIC give rise to~\eqref{eq:P-intro} with $a_1\ne0$,  off-critical quenches correspond to $a_1=0$ in~\eqref{eq:P-intro}, and hence give rise to staircase structure in $n(\lambda)$. In Section~\ref{sec:xxz} we derive analytical predictions for quenches in the XXZ chain. In Section~\ref{sec:numerics} we discuss numerical results supporting the different scenarios. We conclude in Section~\ref{sec:concl}. 

\section{Moment problem for the entanglement spectrum distribution}
\label{sec:theory}

Here we show how to reconstruct the probability density $P(\lambda)$ (cf.~\eqref{eq:P-intro}) describing the lower-part of the entanglement spectrum from the large $\alpha$ expansion of the moments of the reduced density matrix  $M_\alpha$, which are defined as 
\begin{equation}
\label{eq:mom-def}
M_\alpha=\mathrm{Tr}\rho_A^\alpha. 
\end{equation}
Let us consider a subsystem $A$ of length $\ell$ embedded in an infinite system (see Fig.~\ref{fig:cartoon}). 
It is well-known that for both free~\cite{alba2021generalized} and interacting integrable systems~\cite{klobas2022growth} in the hydrodynamic limit $\ell,t\to\infty$, $M_\alpha$ is described as 
\begin{equation}
\label{eq:mom-first}
M_\alpha= e^{-\mathcal{L} F^{}_\alpha}, 
\end{equation}
where  $\mathcal{L}=\ell$ and $\mathcal{L}=2t$ in the long-time and short-time regime, respectively. In~\eqref{eq:mom-first} we neglect contributions that are subleading in the hydrodynamic limit $t,\ell\to\infty$.  The function $F_\alpha$ is different in the two regimes, i.e., at short times or long times, and can be obtained analytically for both free and interacting integrable systems.  The analytical dependence on $\alpha$ is in general quite intricate, which renders challenging to reconstruct $P(\lambda)$ from the moments $M_\alpha$. To proceed, we start with the situation in which $F_\alpha$ admits a series expansion in the limit $\alpha\to\infty$ as 
\begin{equation}
\label{eq:Fa}
F_\alpha=\alpha a_0+\sum_{j=1}^\infty a_j \alpha^{-j}. 
\end{equation}
Again, the coefficients $a_j$ in~\eqref{eq:Fa} depend on the regime that one considers.  It is useful to introduce the parameter $b$ as 
\begin{equation}
b=-\ln(\lambda_{\mathrm{m}}),
\end{equation}
with $\lambda_{\mathrm{m}}$ the largest eigenvalue of $\rho_A$. Since one has $0\le \lambda\le \lambda_\mathrm{m}<1$, in the large $\alpha$ limit the larger eigenvalues dominate the moments $M_\alpha$. As a consequence, it is natural to expect that the $P(\lambda)$ obtained from truncating the expansion of $M_\alpha$ accurately describes the lower part of the entanglement spectrum, i.e., the eigenvalues near $\lambda_\mathrm{m}$. 
Clearly, in the hydrodynamic limit $t,\ell\to\infty$, one has $b=a_0{\mathcal L}$. The  linear scaling of $b$ with $\mathcal{L}$ reflects the linear growth of the R\'enyi entropies at short times, and their volume-law scaling in the steady state. From the exact results for $F_\alpha$, which we discuss in Section~\ref{sec:rule-54},~\ref{sec:ising},~\ref{sec:xxz}, by taking  the 
large $\alpha$ limit, one can determine $a_j$ in~\eqref{eq:Fa}. 

To reconstruct the full distribution $P(\lambda)$ of the 
eigenvalues $\lambda_j$ of $\rho_A$ from  $M_\alpha$, we follow Ref.~\cite{calabrese2008entanglement}, and introduce the generating function $f(z)$ as 
\begin{equation}
\label{eq:f-def}
    f(z) = \frac{1}{\pi} \sum_{\alpha = 1}^{\infty} M_\alpha z^{-\alpha} = \frac{1}{\pi} \int \dd \lambda\, \frac{\lambda\, P(\lambda)}{z - \lambda}.
\end{equation}
Cauchy's theorem implies that one can obtain $P(\lambda)$ 
from $f(z)$ as 
\begin{equation}
\label{eq:P-rep}
    \lambda\, P(\lambda) = \lim_{\epsilon\to 0} \imag \, f(\lambda-\ii\epsilon). 
\end{equation}
Formally, we can rewrite $f(z)$ as 
\begin{equation}
\label{eq:f-1}
f(z)=\frac{1}{\pi}\sum_{\alpha=1}^\infty\sum_{k=0}^\infty \left(\frac{e^{-b}}{z}\right)^\alpha \frac{(-b)^k}{k!}\left(\sum_{j=1}^\infty\frac{a_j}{a_{0}}\alpha^{-j}\right)^k,
\end{equation}
where we expanded in series the exponential in \cref{eq:mom-first} for $M_\alpha$ and we employed the series representation~\eqref{eq:Fa}.
The rightmost term in~\eqref{eq:f-1} can be expanded using the
multinomial theorem. One obtains 
\begin{equation}
f(z)=\frac{1}{\pi}\sum_{\alpha=1}^\infty\sum_{k=0}^\infty \left(\frac{e^{-b}}{z}\right)^\alpha \frac{(-b)^k}{k!}a_0^{-k}\sum_{\sum_j p_j=k} \frac{k!}{\prod_j (p_j!)}\prod_j a_j^{p_j}\alpha^{-\sum_j j p_j}.
\end{equation}
Now the sum over $\alpha$ can be performed exactly, yielding 
\begin{equation}
\label{eq:f-m}
f(z)=\frac{1}{\pi}\sum_{k=0}^\infty \frac{(-b)^k}{k!}a_0^{-k}\sum_{\sum_j p_j=k} \frac{k!}{\prod_j (p_j!)}\prod_j a_j^{p_j} \mathrm{Li}_{\sum_j j p_j}\left(e^{-b}/z\right),
\end{equation}
where $\mathrm{Li}_a(x)$ is the polylogarithmic function. 
We can now use~\eqref{eq:P-rep} and the property 
\begin{equation}
    \lim_{\epsilon\to 0} \imag \Li_{k} (z+\ii\epsilon) = \frac{\pi \lr{\ln z}^{k-1}}{\Gamma(k)} ,
\end{equation}
to obtain 
\begin{multline}
\label{eq:fin-1}
    \lim_{\epsilon\to 0} \imag f(\lambda-\ii\epsilon) 
   =\lambda P(\lambda) =\lambda_{\mathrm{m}}\delta(\lambda-\lambda_{\mathrm{m}}) +\\
    +\sum_{k = 1}^{\infty} \frac{(-a_0)^{-k}}{k!}  \sum_{\sum_j p_j = k} (-\ln\lambda_{\mathrm{m}})^{-\sum_j(j-1) p_j+1}\frac{k!}{\prod_j (p_j!)} \prod_j a_j^{p_j} \frac{\xi^{2\sum_j jp_j-2}}{\Gamma(\sum_j jp_j)}, 
\end{multline}
where the Dirac delta $\delta(\lambda-\lambda_{\mathrm{m}})$ originates from the term with $k=0$ in~\eqref{eq:f-m}.

Let us now discuss some known situations. First, let us consider the case in which we truncate 
the expansion~\eqref{eq:Fa} by keeping only the terms with $j\le 1$. 
Now, \cref{eq:f-1} gives 
\begin{equation}
\label{eq:P-cft}
P(\lambda)=\delta(\lambda-\lambda_\mathrm{m})+\frac{1}{\lambda}\sum_{k=1}^\infty 
(-b)^k\left(\frac{a_1}{a_0}\right)^k\frac{\ln^{k-1}(\lambda_{\mathrm{m}}/\lambda)}{k!\Gamma(k)}.
\end{equation}
The sum in~\eqref{eq:P-cft} can be performed exactly and it gives 
\begin{equation}
\label{eq:PP}
P(\lambda)=\delta(\lambda-\lambda_{\mathrm{m}})+ 
b r_1\frac{I_1\!\lr{2\xi r_1} }{\lambda\xi},\quad \mathrm{with}\,\,\xi:=\sqrt{b\ln(\lambda_{\mathrm{m}}/\lambda)}, 
\end{equation}
where $I_1(x)$ is the modified Bessel function of the first kind, and $r_1=\sqrt{-a_1/a_0}$. 
Now, let us observe that if $M_\alpha$ obey the CFT scaling, 
one has $a_1=-a_0$ (cf.~\cite{calabrese2008entanglement}). 
One can integrate $P(\lambda)$ over $\lambda$ to obtain the cumulative distribution function 
\begin{equation}
\label{eq:nlambda}
n(\lambda)=\int_\lambda^{\lambda_{\mathrm{m}}} d\lambda P(\lambda)=I_0(2r_1\xi),
\end{equation}
with $\xi$ as defined in~\eqref{eq:PP} and $I_0(x)$ the modified Bessel 
function of the first kind. Here $n(\lambda)$ gives the number of eigenvalues of $\rho_A$, which are larger than $\lambda$. Again, in the CFT case $a_1=-a_0$ and $n(\lambda)$ depends only on the central charge of the CFT. 

To check the accuracy of the result for $P(\lambda)$ obtained by truncating the large $\alpha$ expansion of $M_\alpha$, we can 
consider the normalization condition $\int_0^{\lambda_\mathrm{m}} d\lambda \lambda P(\lambda)=1$. From~\eqref{eq:PP} we obtain 
\begin{equation}
\int_{0}^{\lambda_{\mathrm{m}}}\lambda P(\lambda)d\lambda=\lambda_{\mathrm{m}}^{1+a_1/a_0},
\end{equation}
Importantly, only for $a_1=-a_0$, i.e., the CFT case, one has the correct normalization of $P(\lambda)$. Quite generically we have $a_1\ne -a_0$, which implies that  the normalization condition is  violated. Moreover, since $\lambda_\mathrm{m}\simeq e^{-a_0 \mathcal{L}}$, violations become exponentially large at long times. 
From~\eqref{eq:PP} we can also  determine the von Neumann entropy $-\mathrm{Tr}\rho_A\ln(\rho_A)$ as 
\begin{equation}
\label{eq:P-vN}
S_A=-\int_{0}^{\lambda_{\mathrm{m}}}\lambda\ln(\lambda) P(\lambda)d\lambda=-\lambda_{\mathrm{m}}^{1+a_1/a_0}
(1-a_1/a_0)\ln(\lambda_{\mathrm{m}}). 
\end{equation}
Again, for $a_1=-a_0$, i.e., the CFT case, one obtains $S_A=2\ln(\lambda_\mathrm{m})$, i.e., that  the von Neumann entropy is twice the so-called single-copy entanglement~\cite{calabrese2008entanglement}. For $a_1\ne-a_0$, \cref{eq:P-vN} gives a spurious exponential behavior, instead of the expected linear growth with time. 
The failure to capture the normalization condition and  the linear entropy growth (cf.~\cref{eq:P-vN}) signal that there is an exponentially diverging number of  eigenvalues generated by the dynamics that are far from the region $\lambda\approx \lambda_\mathrm{m}$ but are crucial to capture the entanglement dynamics. Still, we anticipate that in the large $\mathcal{L}$ limit  there is an exponentially diverging number of eigenvalues $\lambda\approx \lambda_\mathrm{m}$ that are well described by~\eqref{eq:PP}. Moreover, we will show that at large $\mathcal{L}$ the distribution of the lower part of the entanglement spectrum exhibits scaling behavior when plotted as a function of $\xi$ (cf.~\eqref{eq:PP}). Precisely, the distribution $P(\lambda)$ is given by~\eqref{eq:PP}, which means that in the scaling regime at fixed $\xi$ one can neglect the coefficients $a_j$ with $j>1$ in~\eqref{eq:Fa}, which give subleading contributions at large $\mathcal{L}$. 

Let us now derive the analytic result for the cumulative distribution function $n(\lambda)$. By integrating term by term  in~\eqref{eq:fin-1}  we obtain 
\begin{equation}
\label{eq:n-fin}
    n(\lambda)=   1
    +\sum_{k = 1}^{\infty} \frac{\lr{\ln \lambda_{\mathrm{m}}}^k}{k!} a_0^{-k} \sum_{\sum_j p_j = k} \frac{k!}{\prod_j (p_j!)} \prod_j a_j^{p_j} \frac{\ln^{\sum_j j p_j}\!\lr{\lambda_{\mathrm{m}}/\lambda}}{(\sum_j j p_j)\Gamma(\sum_j jp_j)}, 
\end{equation}
It is useful to rearrange the expansion in~\eqref{eq:n-fin} by expanding in powers of $1/\ln(\lambda_{\mathrm{m}})$.
To lighten the notation, it is convenient to introduce the parameter $s = \sum_{j>1} \lr{j-1} p_j$, in terms of which, at fixed $k$ in~\eqref{eq:n-fin}, one has $\sum_{j\geqslant 1} j p_j = k + s$.
Rewriting $n(\lambda)$ in terms of the scaling variable $\xi = \sqrt{b\ln (\lambda_{\mathrm{m}}/\lambda)}$ one has (using $z\,\Gamma(z) = \Gamma(z+1)$)
\begin{equation}
    n(\lambda) 
    = 1 + \sum_{k = 1}^{\infty} \sum_{\sum_j p_j = k}  \frac{1}{b^s} \lr{\prod_{j>1}\frac{\lr{a_j/a_1}^{p_j}}{p_j!} }  \frac{\lr{-a_1/a_0}^{k}\, \xi^{2\lr{k + s}}}{p_1!\, \Gamma\!\lr{k+s+1}}.
\end{equation}
We can imagine of rearranging the expansion above in powers of $1/b\simeq 1/\mathcal{L}$. As it is clear from the definition of $s$, we notice that $a_j$ with $j>1$ contributes at the lowest order with a term $1/b^{j-1}$. 
This means that higher powers of $1/\alpha$ in the expansion of the moments $M_\alpha$ give rise to higher  powers of $1/b$ in the  distribution $n(\lambda)$ at fixed $\xi$. 
In particular, assuming that the expansion of the momenta includes only odd powers of $1/\alpha$, as will be the case in the following, we notice that at the leading order $\mathcal{O}(b^0)$ only the term $a_1$ contributes.  At the next order $b^{-2}$ one has to include the contribution of $a_3$, whereas  at order $b^{-4}$ one has to include  both $a_3$ and $a_5$. 
Finally, we obtain $n(\lambda)$ as 
\begin{multline}
\label{eq:n-fin-fin}
    n(\lambda)= I_0\!\lr{2\,r_1 \xi } + \frac{1}{b^{2}} \frac{a_3\lr{- a_0}^{1/2}}{a_1^{3/2}}\, \xi^3\, I_{3}\!\lr{2\, r_1 \xi} \\+
      \frac{1}{b^4} \slr{\frac{a_3^2\lr{-a_0}}{2\,a_1^3}\, \xi^6\, I_6\!\lr{2\,r_1 \xi }   +  \frac{a_5\lr{-a_0}^{3/2}}{a_1^{5/2}} \xi^5\, I_5\!\lr{2\,r_1 \xi }  } + \mathcal{O}\!\lr{b^{-6}}.
\end{multline}
Importantly, since $b = - \ln (\lambda_{\mathrm{m}})$ for any fixed value of $\xi$, at sufficiently large $\mathcal{L}$ the cumulative distribution $n(\lambda)$ converges to $n(\lambda) = I_0 \!\lr{2\,r_1\xi}$.

%
\begin{figure}
    \centering
          \includegraphics[width=.8\linewidth]{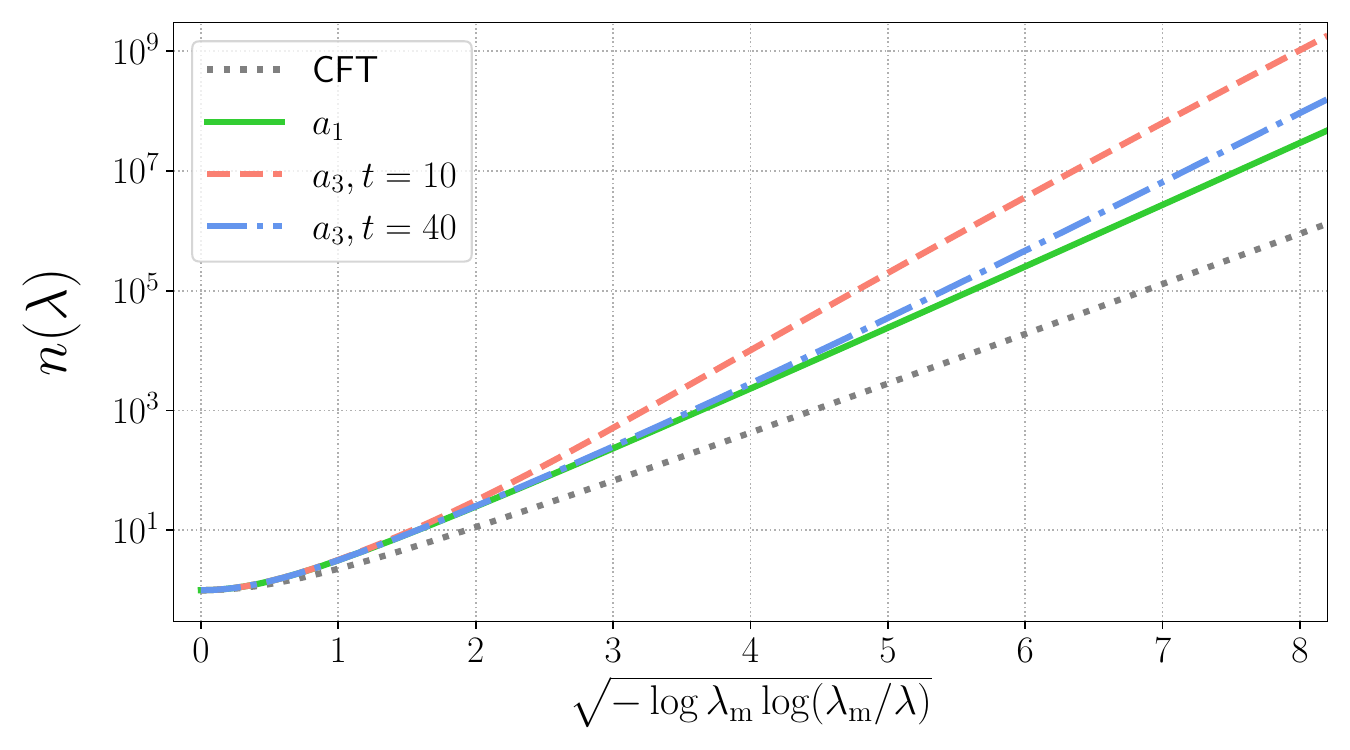}
    \caption{Cumulative distribution function $n(\lambda)$ of the entanglement spectrum plotted versus the scaling variable $\xi=[-\ln(\lambda_\mathrm{m})\ln(\lambda_\mathrm{m}/\lambda)]^{1/2}$, with $\lambda_\mathrm{m}$ the largest eigenvalue of the reduced density matrix. We employ a logarithmic scale on the $y$-axis. The continuous curve is the theoretical result $n(\lambda)=I_0(2r_1\xi)$ (cf.~\eqref{eq:n-fin-fin}) in the limit $t\to\infty$. 
    Here $r_1=\sqrt{-a_1/a_0}$ and we choose $a_0=-0.2, a_1=0.3$. The dashed and dashed dotted lines are the theoretical results at times $t=10, 40$ obtained by including the subleading contributions with $a_3=0.5$ in the expansion~\eqref{eq:Fa}.  At long times one recovers the continuous line. The dotted line is the CFT result, which corresponds to $r_1=1$. 
    }
    \label{fig:theory}
\end{figure} 
%
To illustrate our result, in Fig.~\ref{fig:theory} we 
plot $n(\lambda)$ as  a function of the scaling variable $\xi=\nolinebreak\sqrt{b\ln(\lambda_\mathrm{m}/\lambda)}$. 
We consider an expansion up to the second order in $1/b$, which correspond to retaining only $a_j$ with $j = \clr{0, 1, 3}$.
In particular, here we fix $a_0 = - 0.2, a_1 = 0.3, a_3 \approx 0.5$. 
The dotted line in Fig.~\ref{fig:theory}  is the CFT scaling $n(\lambda)=I_0(2\xi)$. The continuous curve is the CFT-like  prediction~\eqref{eq:n-fin-fin} at 
fixed $\xi$ in the limit $t\to\infty$. In this limit \cref{eq:n-fin-fin} gives 
\begin{equation}
\label{eq:asy-check}
n(\lambda)=I_0(2r_1\xi), 
\end{equation}
with $r_1=\sqrt{-a_1/a_0}$. The dashed and dashed-dotted lines in the figure are the results obtained by including the contribution of $a_3$ in~\eqref{eq:n-fin-fin} for $t=10$ and $t=40$, respectively.  The figure confirms that in the limit $t\to\infty$ one recovers \cref{eq:asy-check}. Finally, it is worth observing that in the limit of large $\xi$, one has 
\begin{equation}
n(\lambda)\simeq \frac{e^{2r_1\xi}}{\sqrt{4\pi r_1\xi}}. 
\end{equation}
Notice that a similar increase with $\xi$ is observed in the ground state entanglement spectrum of systems described by a CFT and gapped ones~\cite{mbeng2017negativity,alba2018entanglementspectrum}. 

\subsection{\texorpdfstring{The ``staircase scenario'' for $a_1=0$}{The staircase scenario for a1=0}}
\label{sec:a1-van}

The results of the previous section were obtained under the assumption that $a_1\ne0$. 
Let us also consider the case in which the moments $F_\alpha$ (cf.~\eqref{eq:Fa}) exhibits a large $\alpha$ expansion of the form 
\begin{equation}
\label{eq:Fa-exp}
F_\alpha=a_0\alpha+\sum_{j=1}^\infty a_je^{-d_j\alpha}, 
\end{equation}
where we neglect exponentially suppressed terms. We anticipate that $F_\alpha$ exhibits the form~\eqref{eq:Fa-exp} for quenches in the rule $54$ chain (see Section~\ref{sec:rule-54}), in the XX chain (see Section~\ref{sec:xxz}), as well as quenches with non-critical TFIC (see Section~\ref{sec:ising}) and in the XXZ chain (see Section~\ref{sec:xxz}). 
We can apply the same strategy as in the previous section defining the generating function $f(z)$ as 
\begin{equation}
f(z)=\frac{1}{\pi}\sum_{\alpha=1}^\infty\sum_{k=0}^\infty 
\left(\frac{e^{-b}}{z}\right)^\alpha \frac{(-b)^k}{k!}
\left(\sum_j\frac{a_j}{a_0}e^{-\alpha d_j}\right)^k.
\end{equation}
Now, the sum over $\alpha$ gives a geometric series as 
\begin{equation}
f(z)=\frac{1}{\pi}\sum_{k=1}^\infty\frac{(-b)^k}{k!}\left(\frac{a_1}{a_0}\right)^k\frac{1}{e^{-d_1 k}(z/\lambda_{\mathrm{m}})-1},
\end{equation}
where we neglected the terms with $d_j$ with $j>1$ because they are exponentially suppressed. One obtains 
\begin{equation}
P(\lambda)=\sum_{k=1}^\infty \frac{(-b)^k}{k!}
\left(\frac{a_1}{a_0}\right)^k\delta(e^{-d_1 k}\lambda-\lambda_\mathrm{m}). 
\end{equation}
The corresponding $n(\lambda)$ is given as 
\begin{equation}
\label{eq:nl-exp}
n(\lambda)=\sum_{k=1}^\infty\frac{(-b)^k}{k!}\left(\frac{a_1}{a_0}\right)^k e^{d_1 k}\Theta(e^{-d_1 k}\lambda-\lambda_\mathrm{m}). 
\end{equation}
Thus, $P(\lambda)$ exhibits a series of delta peaks, which reflect that eigenvalues of $\rho_A$ are organized in clusters of degenerate eigenvalues. As a consequence, $n(\lambda)$ exhibits a ``staircase'' structure with steps at $\lambda=e^{d_1 k}\lambda_\mathrm{m}$, where $k$ is an integer.  

\section{\texorpdfstring{Dynamics of the entanglement spectrum in the rule $54$ chain}{Dynamics of the entanglement spectrum in the rule 54 chain}}
\label{sec:rule-54}

We now apply the results of Section~\ref{sec:theory} to microscopic models. We start considering quenches in the rule $54$ chain. 
The rule $54$ chain was introduced in Ref.~\cite{bobenko1993on}, and it  emerged as a paradigmatic model for interacting integrable dynamics. Several aspects of entanglement  dynamics have been explored, such as dynamics of local operators~\cite{klobas2019time}, operator entanglement dynamics~\cite{gopalakrishnan2018hydro,gopalakrishnan2018operator,alba2019operator,alba2021diffusion,alba2025more}, and entanglement dynamics after quantum quenches~\cite{klobas2021entanglement,klobas2021exact}. The model admits a quantum 
integrable extension~\cite{friedman2019integrable}. 
Importantly, exact results in the rule $54$ chain allowed to conjecture a formula for the growth with time of the R\'enyi entanglement entropies~\cite{klobas2022growth}, generalizing results for their steady-state value~\cite{alba2017quench,alba2017renyi,mestyan2018renyi,alba2019towards} in Bethe ansatz solvable systems. The rule $54$ can be understood as a quantum circuit built out of a $3$-site unitary gate $U$, with matrix elements 
\begin{equation}
\label{eq:54-gate}
U^{s_1's_2's_3'}_{s_1s_2s_3}=\delta_{s_1,s_1'}\delta_{s_2,\chi(s_1,s_2,s_3)}\delta_{s_3,s_3'},
\end{equation}
where $s_j$ take the values $s_j=0,1$, and  the function $\chi$ is defined as 
\begin{equation}
\chi(s_1,s_2,s_3)=s_1+s_2+s_3+s_1s_3\,\mod\,2.
\end{equation}
The time evolution is implemented by applying the gates in two 
distinct time steps as 
\begin{equation}
U=U_o U_e,
\end{equation}
where $U_e$ and $U_o$ are obtained by applying~\eqref{eq:54-gate} on all the even and odd triplets of sites of the chain, respectively, i.e., in a brick-wall fashion. The effect of the gate~\eqref{eq:54-gate} on a triplet $s_1,s_2,s_3$ is to flip the middle spin $s_2$ if either $s_1=1$ or $s_3=1$.  

Let us consider dynamics in the rule $54$ chain after the quench from the state $|\Psi_0\rangle$ defined as~\cite{klobas2022growth} 
\begin{equation}
\label{eq:psi0-54}
|\Psi_0\rangle=\left(\left[\begin{array}{c}1\\0\end{array}\right]\otimes\left[\begin{array}{c}\sqrt{1-\vartheta}\\e^{i\varphi}\sqrt{\vartheta}\end{array}\right]\right)^{\otimes L},
\end{equation}
where $\vartheta,\varphi$ are real parameters. Let us now consider  the dynamics of the R\'enyi entropies $S_\alpha$ in the hydrodynamic limit $t,\ell\to\infty$ at fixed $t/\ell$. Let us first focus on the short-time regime. After the quench from~\eqref{eq:psi0-54} $S_\alpha$ grows linearly with time for any $\alpha$. Remarkably, 
Ref.~\cite{klobas2021exact,klobas2021entanglement} (see also Ref.~\cite{klobas2022growth}) showed that in the 
hydrodynamic regime 
\begin{equation}
S_\alpha(t)=s_\alpha t,
\end{equation}
where $s_\alpha$ is the slope of the linear entanglement growth, and can be determined analytically as 
\begin{equation}
\label{eq:rule-slope}
s_\alpha=\frac{2}{1-\alpha}\ln\left[(1-\vartheta)^\alpha+\frac{\vartheta^\alpha}{y_\alpha}\right].  
\end{equation}
Here $\vartheta$ is the same as in~\eqref{eq:psi0-54}, and $y_\alpha$ is the positive solution of the equation~\cite{klobas2022growth}  
\begin{equation}
\label{eq:eq-y}
\ln y_\alpha=2\ln\left[(1-\vartheta)^\alpha+\frac{\vartheta^\alpha}{y_\alpha}\right]. 
\end{equation}
Notice that \cref{eq:eq-y} admits a closed-form solution as 
\begin{multline}
\label{eq:ya-sol}
y_\alpha= \frac{1}{3} (1-\vartheta )^{2 \alpha}+
\frac{\sqrt[3]{27 \vartheta ^{2 \alpha }+3 \sqrt{3} \sqrt{4 (1-\vartheta )^{3 \alpha } \vartheta ^{3 \alpha }+27 \vartheta ^{4 \alpha }}+18 (1-\vartheta )^{3 \alpha }
   \vartheta ^{\alpha }+2 (1-\vartheta )^{6 \alpha }}}{3 \sqrt[3]{2}}\\
   -\frac{\sqrt[3]{2} \left(-6 (1-\vartheta )^{\alpha } \vartheta ^{\alpha }-(1-\vartheta )^{4 \alpha
   }\right)}{3 \sqrt[3]{27 \vartheta ^{2 \alpha }+3 \sqrt{3} \sqrt{4 (1-\vartheta )^{3 \alpha } \vartheta ^{3 \alpha }+27 \vartheta ^{4 \alpha }}+18 (1-\vartheta )^{3
   \alpha } \vartheta ^{\alpha }+2 (1-\vartheta )^{6 \alpha }}}.
\end{multline}
On the other hand, in the long-time regime $t/\ell\gg1$, one has the volume-law scaling as~\cite{klobas2022growth} 
\begin{equation}
S_\alpha=d_\alpha \ell,
\end{equation}
where $d_\alpha$ reads as~\cite{klobas2022growth} 
\begin{equation}
\label{eq:da}
d_\alpha=\frac{1}{1-\alpha}\ln\left[(1-\vartheta)^\alpha+\vartheta^\alpha\right]. 
\end{equation}
Let us focus on the short-time regime. By substituting \cref{eq:ya-sol} in~\eqref{eq:rule-slope} we obtain $\mathrm{Tr}\rho_A^\alpha$ analytically. 
We do not report the analytic expression because it is cumbersome. 

We can now consider the large $\alpha$ expansion. 
To determine the asymptotic behavior of $\mathrm{Tr}\rho_A^\alpha$, we observe that there is 
a ``critical'' value $\vartheta_c$ at which the behavior of $M_\alpha$ is singular. One straightforwardly obtains that 
\begin{equation}
\label{eq:renyi-crit}
\mathrm{Tr}\rho_A^\alpha\sim \left\{\begin{array}{cc}e^{2t\alpha\ln(1-\vartheta)} & \vartheta<\vartheta_c\\
e^{2/3t\alpha\ln(\vartheta)}&\vartheta\ge \vartheta_c\end{array}\right. 
\end{equation}
with exponentially suppressed corrections in the 
limit $\alpha\to\infty$. For instance, after taking into account  the leading corrections for $\vartheta\ge \vartheta_c$ we obtain  
\begin{equation}
\label{eq:sub-exp}
\ln\mathrm{Tr}\rho_A^\alpha\sim \frac{2t}{3}\ln(z_2)+\frac{2t (1-\vartheta)^\alpha}{3\vartheta^{\alpha/3}}+\frac{2t (1-\vartheta)^{2\alpha}}{18 \vartheta^{2/3\alpha}}+\cdots ,
\end{equation}
where the dots denote subleading terms in the large $\alpha$ limit. 
A similar expansion can be obtained in the region with $\vartheta<\vartheta_c$, although we do not report it. 
The critical $\vartheta_c$ separating the two regimes satisfies the 
condition 
\begin{equation}
\label{eq:54-cft}
\vartheta_c^{1/3}=(1-\vartheta_c),
\end{equation}
which is when the two exponents in~\eqref{eq:renyi-crit} coincide. The critical value is 
\begin{equation}
\label{eq:theta_c}
\vartheta_c\approx 0.317672.
\end{equation}
At $\vartheta=\vartheta_c$ the 
behavior of the moments $M_\alpha$ is simpler. By using the condition~\eqref{eq:54-cft} we obtain 
\begin{equation}
\mathrm{Tr}\rho_A^\alpha=\left[\left(1+\frac{6 \sqrt[3]{47+3 \sqrt{93}}}{14 \sqrt[3]{2}+2 \sqrt[3]{47+3 \sqrt{93}}+\left(94+6 \sqrt{93}\right)^{2/3}}\right) \vartheta_c^{\alpha/3}\right]^{2 \text{t}}.
\end{equation}
Now,  we observe that for $\vartheta=\vartheta_c$, 
$\mathrm{Tr}\rho_A^\alpha$ exhibits a purely exponential decay with $t$. This implies that only $a_0$ in~\eqref{eq:Fa} is nonzero, and all the $a_j$ with $j\ge1$ are zero. For $\vartheta\ne\vartheta_c$ subleading exponential corrections are present (cf. for instance~\eqref{eq:sub-exp}). As we showed in Section~\ref{sec:cft-like}, the presence of several exponential terms in the large $\alpha$ expansion of $M_\alpha$ implies a ``staircase'' structure in the cumulative distribution $n(\lambda)$ of the  entanglement spectrum. On the other hand, for $\vartheta=\vartheta_c$ a single step in $n(\lambda)$ is expected. Finally, let us mention that a similar behavior is present in the long-time regime. Indeed, by expanding~\eqref{eq:da} for large $\alpha$ one obtains a sum of exponentially decaying terms, which implies the presence of a staircase in $n(\lambda)$. This does not happen at $\vartheta=1/2$, where one has a single exponential term, and hence a single step in $n(\lambda)$. It is interesting to observe that the singular $\vartheta_c$ at which the structure of the entanglement spectrum changes is given by~\eqref{eq:theta_c} in the short-time regime, whereas it is $\vartheta_c=1/2$ in the long-time one. 

\section{Magnetic-field quench in transverse field Ising chain (TFIC)}\label{sec:ising}

We now apply the framework of Section~\ref{sec:theory} to magnetic-field quenches in the TFIC. The model is defined by the Hamiltonian as 
\begin{equation}\label{eq:isingHam}
    H = -J\sum_{i=1}^L \left[\px_i \px_{i+1} + h \pz\right],
\end{equation}
where $J,h$ are real parameters, and $\sigma_i^{x,z}$ Pauli matrices acting at site $i$. We employ periodic boundary conditions identifying sites $1$ and $L+1$. 
We also set $J=1$. 
After a Jordan-Wigner transformation \cref{eq:isingHam} is mapped onto a quadratic fermionic system. Notice that there are some subtleties in the mapping between the  spins and the fermions, due to the fermion parity, which affect   the boundary conditions that one has to employ for the fermion model~\cite{calabrese2012quantum}. We ignore these subtleties because they would lead to subleading corrections in the hydrodynamic limit, and work directly with the fermionic model with periodic boundary conditions. 
The model is diagonalized by a combination of Fourier 
transform and Bogoliubov transformation. \cref{eq:isingHam} becomes 
\begin{equation}
H=\sum_k \varepsilon_k b^\dagger_kb_k, 
\end{equation}
with $b_k$ Bogoliubov modes, which are linear combination of the lattice fermions~\cite{calabrese2012quantum,alba2021generalized}, and $k$ the momentum. The energy dispersion relation is 
\begin{equation}
    \varepsilon_k =  \sqrt{1 + h^2 - 2 h \cos k}. 
\end{equation}
In the thermodynamic limit, the ground state of~\eqref{eq:isingHam} exhibits a paramagnetic phase for $h>1$, and a ferromagnetic one for $h<1$. The two phases are separated by a quantum phase transition at $h=h_c=1$. 
The presence of the critical point is reflected in cusp-like singularity at $k=0$ in the dispersion relation $\varepsilon(k)$. 
In the thermodynamic limit, the excitations have group velocity $v_k$ as 
\begin{equation}
    v_k = \frac{\partial \varepsilon_k}{\partial k} = \frac{2 h \sin k}{\sqrt{1+h^2-2h \cos k}}. 
\end{equation}
In our protocol, at time $t = 0$ the system is prepared in the ground state of the Ising Hamiltonian~\eqref{eq:isingHam} with transverse magnetic field $h_0$. At time $t>0$ we quench the magnetic field to $h$. 
It is convenient to define  Bogoliubov angle $\Delta_k$ that parametrizes the quench as 
\begin{equation}
\label{eq:Dk}
    \Delta_k = \frac{1 + h h_0-(h+h_0) \cos k}{\sqrt{1+h^2-2h\cos k} \, \sqrt{1+h_0^2-2h_0\cos k}}.
\end{equation}
Importantly, the density of Bogoliubov modes $n_k$ is given as 
\begin{equation}
\label{eq:density}
    n_k =\langle b_k^\dagger b_k\rangle= \frac{1-\Delta_k}{2}, 
\end{equation}
and is preserved during the dynamics. 

Within the quasiparticle picture for entanglement spreading in free-fermion and free-boson models the dynamics of the R\'enyi entropies of a subsystem $A$ of length $\ell$ embedded in an infinite chain (see Fig.~\ref{fig:cartoon}) is given as~\cite{fagotti2008evolution,alba2017entanglement,alba2018entanglement} 
\begin{equation}
\label{eq:quasi}
    S_A^{(\alpha)} = \int_{-\pi}^\pi \frac{\dd k}{2\pi}\, s^{(\alpha)}_k\, \min\!\lr{2\abs{v_k}t , \ell}. 
\end{equation}
\cref{eq:quasi} holds in the hydrodynamic limit 
$t,\ell\to\infty$ with the ratio $t/\ell$ arbitrary but fixed. \cref{eq:quasi} admits a quasiparticle picture because $\min(|v_k|t,\ell)$ is the number of entangled pairs of quasiparticles that are produced after the quench, and that at time $t$ are shared between $A$ and the rest. Crucially, the quasiparticle picture interpretation breaks down when considering the dynamics of R\'enyi entropies in interacting integrable systems~\cite{klobas2022growth}. 
In~\eqref{eq:quasi}, $s^{(\alpha)}_k$ is the contribution of the quasiparticles to the R\'enyi entropies, and is  given as 
\begin{equation}
    s^{(\alpha)}_k = \frac{1}{1-\alpha}\, \ln \slr{n_k^\alpha + \lr{1-n_k}^\alpha}, 
\end{equation}
where $n_k$ is the given in~\eqref{eq:density}. 
To proceed, it is convenient to rewrite \cref{eq:quasi} as 
\begin{multline}
\label{eq:ln-tr}
    \ln (\Tr \rho_A^\alpha) = \int_{-\pi}^{\pi} \frac{\dd k}{2\pi}\, \ln\slr{\lr{\frac{1+\Delta_k}{2}}^\alpha + \lr{\frac{1-\Delta_k}{2}}^\alpha} \min\!\lr{\abs{v_k}t , \ell}\\
    =\int_{-\pi}^{\pi} \frac{\dd k}{2\pi}\, \clr{\alpha\ln\slr{\frac{1+\Delta_k}{2}} + \ln\slr{1+\lr{\frac{1-\Delta_k}{1+\Delta_k}}^\alpha} }\min\!\lr{\abs{v_k}t , \ell}.
\end{multline}
In the following, we will consider separately the short time regime, in which $|v_k| t \ll \ell$ for all $k \in \slr{-\pi, \pi}$, and the long time one, with $|v_k| t \gg \ell$. In both cases, we consider the large $\alpha$ expansion of the R\'enyi entropies, which allows us to describe the lower part of the entanglement spectrum. 
In the two regimes from~\eqref{eq:quasi} we obtain 
\begin{equation}
\label{eq:mom-quench}
M_\alpha=\mathrm{Tr}\rho_A^\alpha=\left\{
\begin{array}{ll}
 \exp\left[2t(1-\alpha)\int_{-\pi}^\pi\frac{dk}{2\pi}|v_k|s_\alpha(k)\right]& 2v_{\max}t/\ell\ll 1\\
 \exp\left[\ell(1-\alpha)\int_{-\pi}^\pi\frac{dk}{2\pi}s_\alpha(k)\right] & 2v_{\max}t/\ell\gg1\end{array}
\right.,
\end{equation}
where $v_\mathrm{max}$ is the maximum group velocity. 
It is important to stress that  we derived \cref{eq:mom-quench} by using the result of the quasiparticle picture~\eqref{eq:quasi}. In doing that we are neglecting subleading corrections that vanish in the hydrodynamic limit.

\subsection{\texorpdfstring{Large-$\alpha$ expansion in the long-time regime}{Large-alpha expansion in the long-time regime}}
\label{sec:long}

Here we determine the entanglement spectrum distribution in the long-time limit, 
i.e., in the steady state after the quench. We are interested in the large $\alpha$ limit of the moments $M_\alpha=\mathrm{Tr}\rho_A^\alpha$. We consider quenches $h_0\to h$, with $h=1$, i.e., quenches in which the final Hamiltonian is critical.  This leads to a CFT-like structure in the entanglement spectrum (see Section~\ref{sec:theory}). We comment at the end on the case with $h\ne 1$, i.e., the non-critical case. 

To expand the moments $M_\alpha$ in powers of $1/\alpha$, we use the fact that (cf.~\eqref{eq:ln-tr}) for $h=1$ in the large $\alpha$ limit we have  
\begin{equation}
\label{eq:cond}
    \lim_{\alpha\to\infty} \lr{\frac{1-\Delta_k}{1+\Delta_k}}^\alpha = \begin{cases}
        1,  &k = 0,\\
        0,  &\text{otherwise,}
    \end{cases}
\end{equation}
where $\Delta_k$ is defined in~\eqref{eq:Dk}. 
We need to compute an integral of the form 
\begin{equation}
    \int_{-\pi}^\pi \frac{\dd k}{2 \pi} \, \ln\!\slr{ 1 + e^{- \alpha f(k)} }, 
\end{equation}
where we can Taylor expand the function $f(k)$ around $k=0$ because of~\eqref{eq:cond}.  We obtain  
\begin{equation}
    f(k) = \sum_{n = 1}^{\infty} b_n \abs{k}^n, \qquad \text{with } b_1 >0.
\end{equation}
One can verify that  $b_{2n}=0$. We have 
\begin{multline}
    \int_{-\pi}^{\pi} \frac{\dd k}{2 \pi} \, \ln\!\slr{ 1 + e^{- \alpha f(k)} } = 
     \frac{1}{2\pi} \int_{-\alpha\pi}^{\alpha\pi} \frac{\dd x}{\alpha} \, \ln\!\slr{ 1 + e^{- b_1 \abs{x}} \exp\!\clr{- \sum_{n=1}^\infty b_{2n+1} \frac{\abs{x}^{2n+1}}{\alpha^{2n}}} }.
\end{multline}
We can now exploit that the integrand is symmetric around $x=0$, and we can expand it in the large $\alpha$ limit. After integrating each term of the expansion, we obtain 
\begin{equation}
    \int_{-\pi}^{\pi} \frac{\dd k}{2 \pi} \, \ln\!\slr{ 1 + e^{- \alpha f(k)} } = \frac{\pi}{12\, b_1}\frac{1}{\alpha} - \frac{7\pi^3\,b_3}{120\,b_1^4} \frac{1}{\alpha^3} - \frac{31 \pi^5 \lr{b_1b_5 - b_3^2}}{252\, b_1^7} \frac{1}{\alpha^5} + \mathcal{O}\!\lr{\frac{1}{\alpha^7}}.
\end{equation}
For the magnetic field quench in the Ising chain with $h=1$, we obtain 
\begin{multline}
\label{eq:ln-tr-l}
    \ln\Tr\rho_A^\alpha = -\alpha\ell \slr{\ln 2 - \frac{1}{2\pi}\int_{-\pi}^{\pi} \dd k\, \ln \!\lr{1 + \sgn\!\lr{1+h_0} \Delta_k} }  
    + \ell\,\frac{\pi}{12} \abs{\frac{h_0-1}{h_0+1}} \left[\frac{1}{\alpha}\right. + \\\left.- \frac{1}{\alpha^3} \frac{7\pi^3}{240} \frac{\lr{1-6h_0 +h_0^2}}{\lr{h_0+1}^2} + \frac{1}{\alpha^5}\frac{31 \pi^5}{40320}\frac{\lr{5 - 92 h_0 + 190 h_0^2 - 92 h_0^3 + 5 h_0^4}}{\lr{h_0+1}^4} + \mathcal{O}\!\lr{\frac{1}{\alpha^7}}\right],
\end{multline}
which is of the form~\eqref{eq:Fa}. From~\eqref{eq:ln-tr-l} one can extract the coefficients $a_j$ of \cref{eq:Fa}. One should observe that for quenches with $h\ne1$ \cref{eq:cond} is violated. This implies that  only $a_0$ in~\eqref{eq:Fa} is nonzero, the remaining terms being exponentially suppressed in the large $\alpha$ limit.

\subsection{\texorpdfstring{Large-$\alpha$ expansion in the short-time regime}{Large-alpha expansion in the short-time regime}}
\label{sec:short}

At short times $t < \ell/(2v_{\mathrm{max}})$,  the R\'enyi entropies grow 
linearly with time. The strategy to perform the large $\alpha$ analysis is similar to Section~\ref{sec:long}. 
In the large $\alpha$ limit, from~\eqref{eq:ln-tr}  one obtains that 
\begin{multline}
\label{eq:rate-l-alpha}
    \ln \Tr \rho_A^\alpha = 2t \int_{-\pi}^{\pi}\frac{\dd k}{2\pi}\, \abs{v_k} \ln\slr{\lr{\frac{1+\Delta_k}{2}}^\alpha + \lr{\frac{1-\Delta_k}{2}}^\alpha}= \\
     - 2t\alpha \slr{\frac{2}{\pi}\ln 2 - \frac{1}{2\pi}\int_{-\pi}^{\pi} \dd k\,\abs{v_k}  \ln \!\lr{1 + \sgn\!\lr{1+h_0} \Delta_k}} + 
     2t\, \frac{\pi}{12}\abs{\frac{h_0 - 1}{h_0 + 1}} \left[\frac{1}{\alpha}\right. +\\ \left.- \frac{1}{\alpha^3} \frac{7\pi^3}{120} \frac{\lr{1-4 h_0 +h_0^2}}{\lr{h_0+1}^2} + \frac{1}{\alpha^5} \frac{31\pi^5}{2520} \frac{\lr{1-11 h_0 +21h_0^2 -11h_0^3 + h_0^4} }{\lr{h_0+1}^4}  + \mathcal{O}\left(\!{\frac{1}{\alpha^7}}\right)\right]. 
\end{multline}
It is interesting to compare the large $\alpha$ expansion of~\eqref{eq:rate-l-alpha} and~\eqref{eq:ln-tr-l}. In both cases the expansion is of the form~\eqref{eq:Fa}. However, the leading term  $a_0$, i.e., the prefactor of the $\alpha$ term in~\eqref{eq:rate-l-alpha} and~\eqref{eq:ln-tr-l} is different, and it depends on the full dispersion  of the model, as it is clear from the integration over the full Brillouin zone in~\eqref{eq:rate-l-alpha} and~\eqref{eq:ln-tr-l}. On the other hand, it is remarkable that the coefficient $a_1$ of~\eqref{eq:rate-l-alpha} is obtained from that in~\eqref{eq:ln-tr-l} by replacing $\ell\to2t$. This reflects a ``duality'' between space and time in the distribution of the entanglement spectrum. This is particularly interesting because the cumulative distribution $n(\lambda)$ depends only on $a_1$. Moreover, $a_1$ is obtained from the expansion of $\Delta_k$ around $k=0$, and not on the full functional form of $\Delta_k$. Finally, it is intriguing that there exist a value of $h_0$ for which $a_0=-a_1$. From~\eqref{eq:ln-tr-l} and~\eqref{eq:rate-l-alpha} it is straightforward to show that this corresponds to $h^*_0\approx0.38175$. At $h_0^*$ the distribution of the lower part of the entanglement spectrum in the scaling limit is the same as that of a CFT. 

\section{Dynamics of the entanglement spectrum after quenches in the XXZ chain}
\label{sec:xxz}

To address the validity of the results of Section~\ref{sec:theory} in Bethe ansatz solvable systems,  
here we consider the spin-$1/2$ XXZ chain, which is a paradigmatic integrable spin system, and it is described by the Hamiltonian  
\begin{equation}
\label{eq:xxz-ham}
H=\frac{J}{4}\sum_{j=1}^L[\sigma_j^x\sigma_{j+1}^x+\sigma_j^y\sigma_{j+1}^y+\Delta(\sigma_j^z\sigma_{j+1}^z-1)], 
\end{equation}
where $J,\Delta$ are real parameters, $\sigma_j^{x,y,z}$ are the Pauli matrices, and we use periodic boundary conditions. 
Here we set $J=1$ and restrict ourselves to $\Delta>1$. The XXZ spin chain is solvable by the Bethe ansatz~\cite{takahashi1999thermodynamics}. We consider quantum quenches from the N\'eel state $|\Psi_N\rangle$ and the Majumdhar-Ghosh state $|\Psi_{MG}\rangle$, which are defined as 
\begin{align}
\label{eq:neel}
& |\Psi_N\rangle=\frac{1}{\sqrt{2}}\left(|\uparrow\downarrow\rangle^{\otimes L/2}+|\downarrow\uparrow\rangle^{\otimes L/2}\right)\\
\label{eq:mg-state}
& |\Psi_{MG}\rangle=\frac{1}{\sqrt{2}}\left(|\uparrow\downarrow\rangle-|\downarrow\uparrow\rangle\right)^{\otimes L/2}.
\end{align}
Both these states are integrable~\cite{piroli2017what}, which means that the steady state arising after the quench can be characterized analytically~\cite{caux2016the,pozsgay2018overlaps}. 
More in general the steady-state arising after quenches from generic low-entangled initial states can be also obtained, at least numerically~\cite{ilievski2016string}. The dynamics of the von Neumann entropy after the quenches from~\eqref{eq:neel} and~\eqref{eq:mg-state} is described by the quasiparticle picture~\cite{alba2017entanglement,alba2018entanglement}. The study of the dynamics of the R\'enyi entropies in interacting integrable systems was started in Ref.~\cite{alba2017quench}, which derived the steady-state value of the R\'enyi entropies of a finite subsystem embedded in an infinite chain in the limit of large subsystem size $\ell\to\infty$. The intermediate-time dynamics of the R\'enyi entropies is a challenging problem. In Ref.~\cite{klobas2022growth} it was shown that although the R\'enyi entropies grow linearly with time at short times $t/\ell\ll 1$, they defy the quasiparticle picture interpretation, meaning that the growth cannot be interpreted in terms of shared entangled pairs of quasiparticles. Still, by generalizing exact results available for quenches in the rule $54$ chain, Ref.~\cite{klobas2022growth} conjectured a formula for the growth rate of the R\'enyi entropies with time.

\subsection{R\'enyi entropies in the  short-time regime: Bethe ansatz results}
\label{sec:renyi-bethe-a}

Here we characterize the dynamics of the lower part of the entanglement spectrum in the short-time regime at $t/\ell\ll 1$. 
The starting point is the growth rate of the R\'enyi entropies, which is obtained  from the thermodynamic macrostate, i.e., the statistical ensemble, describing local steady-state properties. In the following we briefly overview the computation of the growth rate of R\'enyi entropies (see Ref.~\cite{klobas2022growth} for the details). A generic integrable system is characterized by the presence of stable families of \emph{quasiparticles}. Each quasiparticle is identified by a real parameter $\lambda$, which is called rapidity. For models mappable to free-fermions, such as the TFIC, $\lambda$ is the quasimomentum. Moreover, generic interacting integrable systems feature several species of different quasiparticles, which are typically bound states of elementary ones, and are labeled by the number $n$ of quasiparticles forming the bound state. Thus, quasiparticles are identified by the rapidity and the species index as $(\lambda,n)$. Here we focus on the XXZ chain at $\Delta>1$, for which $\lambda\in[-\pi/2,\pi/2]$ and $n$ can take any value in $[1,\infty)$, i.e., bound states of arbitrary sizes are allowed. Any legit \emph{thermodynamic} macrostate for the chain is fully characterized by the rapidity densities 
$\rho_n(\lambda)$ and by the hole densities $\rho^h_{n}(\lambda)$, i.e., the density of unoccupied rapidities. The total density $\rho^t_{n}(\lambda)=\rho_n(\lambda)+\rho^h_{n}(\lambda)$ is also nontrivial, because the quasiparticles are interacting, unlike in free models. The densities $\rho_n(\lambda),\rho_n^h(\lambda),\rho^t_n(\lambda)$ are not independent but satisfy a set of nonlinear coupled integral equations, which are the thermodynamic version of the famous Bethe equations~\cite{takahashi1999thermodynamics}. We do not report them, as they are not needed here. Let us introduce the densities $\eta_n(\lambda)$ and $\vartheta_n(\lambda)$ as 
\begin{equation}
\label{eq:eta-theta}
\eta_n(\lambda)=\frac{\rho^h_n(\lambda)}{\rho_n(\lambda)},\quad \vartheta_n=\frac{1}{1+\eta_n(\lambda)}. 
\end{equation}
Here $\vartheta_n(\lambda)=\rho_n(\lambda)/\rho_n^t(\lambda)$ are the so-called filling functions.  Notice that the filling functions $\vartheta_n(\lambda)$ play the same role as the $\vartheta$ (cf.~\eqref{eq:rule-slope}) for the rule $54$, although they depend on the species index $n$ and on the rapidity $\lambda$. The filling functions, together with the Thermodynamic Bethe Ansatz (TBA) equations uniquely identify the generic thermodynamic macrostate of the system. For the quenches from~\eqref{eq:neel} and~\eqref{eq:mg-state} the $\vartheta_n(\lambda)$ are known analytically (see, for instance, Ref.~\cite{klobas2022growth}). 

Determining the behavior of the R\'enyi entropy requires minimal knowledge about the microscopic model. Precisely, one has to know the bare energies  $\varepsilon_n(\lambda)$ and momenta $p_n(\lambda)$  of the quasiparticles, which are known for any Bethe ansatz integrable system. In particular, for the XXZ chain, one has~\cite{takahashi1999thermodynamics} 
\begin{align}
& \frac{\varepsilon'_n(\lambda)}{2\pi}=\frac{1}{\pi}\frac{\sin(2\lambda)\sinh(\eta)\sinh(n\eta)}{[\cosh(n\eta)-\sin(2\lambda)]^2}\\
& \frac{p'_n(\lambda)}{2\pi}=\frac{1}{\pi}\frac{\sinh(n\eta)}{\cosh(n\eta)-\sin(2\lambda)}, 
\end{align}
where $\eta=\mathrm{arccosh}(\Delta)$, and the prime in $p_n'(\lambda),\varepsilon_n'(\lambda)$ denotes the derivative with respect to $\lambda$. 
Furthermore, another crucial ingredient is the scattering kernel $T_{nm}(\lambda,\mu)$, which encodes the interactions between quasiparticles. 
For the XXZ chain the kernel $T_{nm}(\lambda,\mu)$ 
is of the difference form as $T_{nm}(\lambda-\mu)$, and it is given as 
\begin{align}
& T_{nn}(\lambda)=\frac{1}{2\pi}\sum_{k=1}^n p'_{2k}(\lambda)\\
& T_{nm}(\lambda)=\frac{1}{2\pi}\sum_{k=0}^{\frac{n+m-|n-m|}{2}}p'_{|n-m|+2k}(\lambda). 
\end{align}
For the following, it is useful to introduce the 
notation $T({\pmb{\lambda},\pmb{\mu}})=T_{nm}(\lambda,\mu)$, and 
\begin{equation}
\label{eq:int-sum}
\int d\pmb{\lambda}=\sum_{n=1}^\infty\int_{-\pi/2}^{\pi/2} d\lambda, 
\end{equation}
and $\int_+$ is the same as~\eqref{eq:int-sum} with the integration restricted to positive rapidity. 

We are now ready to discuss the prediction for the slope of the growth of the R\'enyi entropies. Let us consider the half-system R\'enyi entropy $S_\alpha(t)$ in an infinite chain after a global quantum quench. Since subsystem $A$ is infinite, for typical global quenches $S_\alpha$ grows linearly with time. We define the slope $s_\alpha$ of the growth as 
\begin{equation}
s_\alpha=\lim_{t\to\infty} \frac{S_\alpha(t)}{t}. 
\end{equation}
The slope $s_\alpha$ of the R\'enyi entropy growth was conjectured in Ref.~\cite{klobas2022growth}, and it is given as 
\begin{equation}
\label{eq:sa-klobas}
s_\alpha=\frac{2}{1-\alpha}\int_+d\pmb{\mu}\frac{{\varepsilon}'(\pmb{\mu})}{2\pi}
\ln\left[(1-\vartheta(\pmb{\mu}))^\alpha+\frac{\vartheta^\alpha(\pmb{\mu}))}{y_\alpha(\pmb{\mu})}\right]
\end{equation}
where the function $y_\alpha(\pmb{\lambda}):=y_{\alpha,n}(\lambda)$ satisfies the integral equation as 
\begin{equation}
\label{eq:y-eq}
\ln y_\alpha(\pmb{\lambda})=\int_+ d\pmb{\mu}\left[T(\pmb{\mu},\pmb{\lambda})-T(\pmb{\mu},-\pmb{\lambda})\right] 
\ln\left[(1-\vartheta(\pmb{\mu}))^\alpha+\frac{\vartheta^\alpha(\pmb{\mu})}{y_\alpha(\pmb{\mu})}\right], 
\end{equation}
and $\vartheta(\pmb{\lambda})$ are the filling functions that identify the macrostate describing the  steady state. 
Notice that \cref{eq:y-eq} is an infinite system of coupled integral equations, one for each bound state. 
It is possible to rewrite~\eqref{eq:y-eq} in a partially decoupled form as 
\begin{equation}
\label{eq:decoup}
\ln y_{\alpha,n}(\lambda)=s\star\mathrm{sgn}(\cdot)[(1-\vartheta_{n-1})^\alpha y_{\alpha,n-1}^{\mathrm{sgn}(\cdot)}+\vartheta_{n-1}^{\alpha}](\lambda)+
s\star\mathrm{sgn}(\cdot)[(1-\vartheta_{n+1})^\alpha y_{\alpha,n+1}^{\mathrm{sgn}(\cdot)}+\vartheta_{n+1}^{\alpha}](\lambda)
\end{equation}
with 
\begin{equation}
\label{eq:decoup-2}
\ln y_{\alpha,1}(\lambda)=s\star\mathrm{sgn}(\cdot)[(1-\vartheta_{2})^\alpha y^{\mathrm{sgn}(\cdot)}_{\alpha,2}+\vartheta_{2}^{\alpha}](\lambda),
\end{equation}
where $s(\lambda)$ is defined as 
\begin{equation}
s(\lambda)=\frac{1}{2\pi}\sum_{k=-\infty}^\infty \frac{e^{-2i k\lambda}}{\cos(k\eta)}, 
\end{equation}
and $\mathrm{sgn}(\cdot)$ is the sign function, with $\mathrm{sgn}(0)=0$. In~\eqref{eq:decoup} and~\eqref{eq:decoup-2} one has to use that $\ln y_{\alpha,n}(\lambda)$ is an odd function of $\lambda$. The $\star$ symbol denotes convolution  as 
\begin{equation}
(f\star g)(\lambda):=\int_{-\pi/2}^{\pi/2}d\mu f(\lambda-\mu)g(\mu).  
\end{equation}
We can rewrite the equations more explicitly as 
\begin{multline}
\label{eq:decoup-x}
\ln y_{\alpha,n}(\lambda)=s_+\star[(1-\vartheta_{n-1})^\alpha y_{\alpha,n-1}+\vartheta_{n-1}^{\alpha}](\lambda)-
s_-\star[(1-\vartheta_{n-1})^\alpha y_{\alpha,n-1}+\vartheta_{n-1}^{\alpha}](\lambda)\\+
s_+\star[(1-\vartheta_{n+1})^\alpha y_{\alpha,n+1}+\vartheta_{n+1}^{\alpha}](\lambda)-
s_-\star[(1-\vartheta_{n+1})^\alpha y_{\alpha,n+1}+\vartheta_{n+1}^{\alpha}](\lambda)
\end{multline}
with 
\begin{equation}
\label{eq:decoup-2x}
\ln y_{\alpha,1}(\lambda)=s_+\star[(1-\vartheta_{2})^\alpha y_{\alpha,2}+\vartheta_{2}^{\alpha}](\lambda)- 
s_-\star[(1-\vartheta_{2})^\alpha y_{\alpha,2}+\vartheta_{2}^{\alpha}](\lambda)
\end{equation}
where we introduced $s_\pm$ such that 
\begin{equation}
s_\pm\star f:=\int_0^{\pi/2} s(\lambda\mp\mu)f(\mu). 
\end{equation}
Again, for the quenches from~\eqref{eq:neel} and~\eqref{eq:mg-state}, the filling functions $\vartheta(\pmb{\lambda})$ are known analytically~\cite{klobas2022growth}. 

\subsection{\texorpdfstring{Large-$\alpha$ expansion of the R\'enyi entropies}{Large-alpha expansion of the R\'enyi entropies}}
\label{sec:la-xxz}

As for the quenches in the TFIC and the rule $54$ chain, the strategy to obtain 
the distribution of the entanglement spectrum is to consider the large $\alpha$ expansion of $S_\alpha$. 
Let us start by  determining the large $\alpha$ expansion of  
$y_{\alpha,n}(\lambda)$. We employ the ansatz for $y_{\alpha,n}(\lambda)$ (cf.~\eqref{eq:y-eq}) as 
\begin{equation}
	y_{\alpha,n}=e^{\alpha \gamma_n(\lambda)},
\end{equation}
where $\gamma_n(\lambda)$ is to be determined. 
It is straightforward to obtain the leading order in the 
large $\alpha$ limit of the partially decoupled equations~\eqref{eq:decoup-2x} as  
\begin{multline}
\label{eq:large-alpha-d}
\gamma_n(\lambda)=(s_+-s_-)\star\Big\{\Theta[(1-\vartheta_{n-1})e^{\gamma_{n-1}}-\vartheta_{n-1}]\big(\ln(1-\vartheta_{n-1})+\gamma_{n-1}\big)
\\+\Theta[\vartheta_{n-1}-(1-\vartheta_{n-1})e^{\gamma_{n-1}}]\ln \vartheta_{n-1}
\Big\}+n\to n+2. 
\end{multline}
In the equations~\eqref{eq:large-alpha-d} only linear terms in $\gamma_n$ appear, except for the Heaviside theta functions $\Theta(x)$. 
This prevents from obtaining an exact solution. 
Alternatively, similar manipulations allow us to rewrite the 
coupled equations~\eqref{eq:y-eq} for $\gamma_n(\pmb{\lambda})$ as 
\begin{multline}
\label{eq:large-alpha-c}
\gamma(\pmb{\lambda})=\int_+d\pmb{\mu}[T(\pmb{\mu},\pmb{\lambda})-T(\pmb{\mu},-\pmb{\lambda})]
\Big\{\Theta((1-\vartheta(\pmb{\mu}))e^{\gamma(\pmb{\mu})}-\vartheta(\pmb{\mu}))\ln(1-\vartheta(\pmb{\mu}))\\+
\Theta(\vartheta(\pmb{\mu})-(1-\vartheta(\pmb{\mu}))e^{\gamma(\pmb{\mu})})\big[\ln\vartheta(\pmb{\mu})-\gamma(\pmb{\mu})\big]
\Big\}. 
\end{multline}
%
%
\begin{figure}
    \centering
          \includegraphics[width=.8\linewidth]{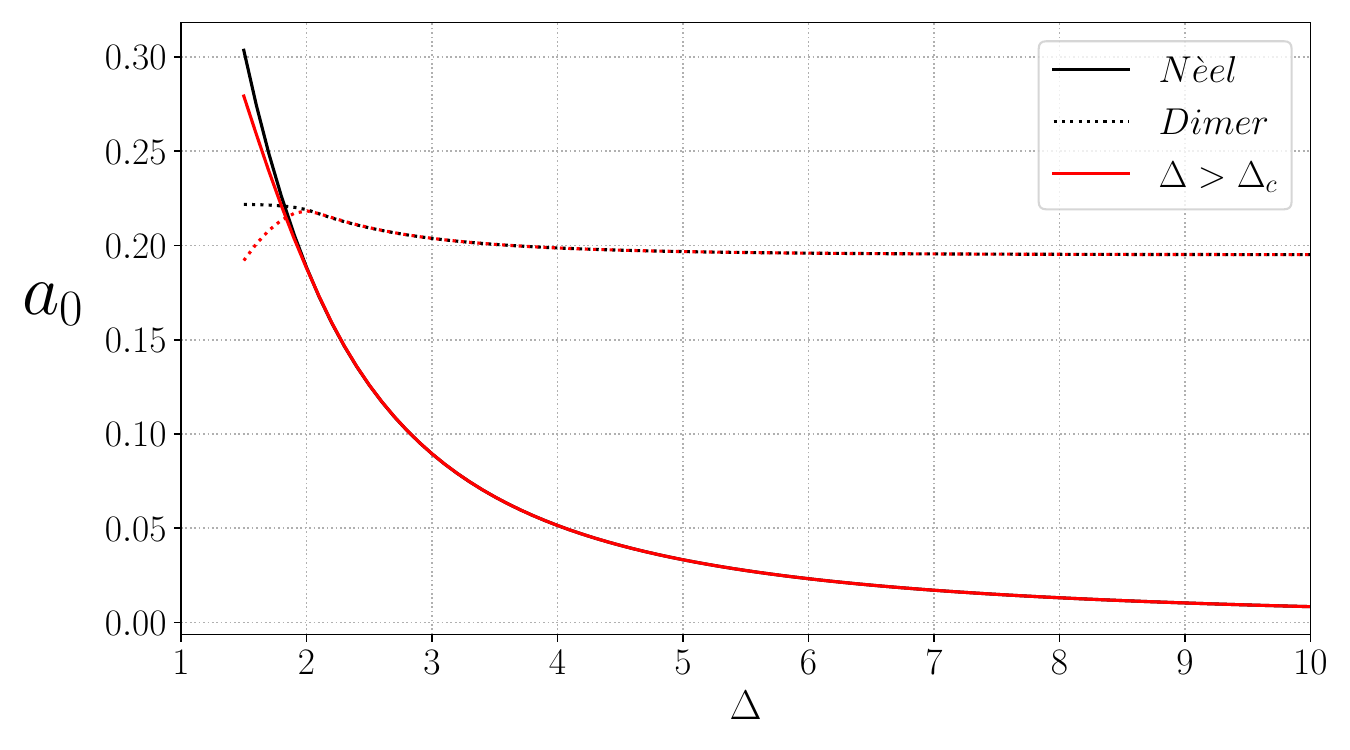}
    \caption{Quantum quenches in the XXZ chain. The coefficient $a_0$ (cf.~\eqref{eq:Fa}) plotted as a function of the chain anisotropy $\Delta$. The full line is the exact result for the N\'eel quench, whereas the dotted line is for the dimer quench. The red lines are obtained from~\eqref{eq:large-alpha-sol} and~\eqref{eq:large-alpha-sol-dimer} which are valid for $\Delta>\Delta_c$. 
    }
    \label{fig:a0-xxz}
\end{figure} 
%
In some regimes it is possible to obtain $\gamma_n(\lambda)$ analytically in terms 
of $\vartheta_n(\lambda)$. To this purpose it is useful to consider the large $\Delta$ limit, 
focusing  on the quench from the N\'eel state. One can verify by using the exact results for $\vartheta_n(\lambda)$ for the N\'eel quench~\cite{klobas2022growth} that $\vartheta_n(\lambda)\to 1$ 
for odd $n$ and for any $\lambda$. This implies that only the second term 
in the integral in~\eqref{eq:large-alpha-d} contributes 
in the equation for $\gamma_n$ with $n$ odd in the left-hand side. 
Moreover, for even $n$, the first term 
in~\eqref{eq:large-alpha-d} contributes only for $\lambda<\lambda_{n\pm1}^\star$ 
(for $\Delta\to\infty$ one has $\lambda^\star_{n\pm1}=\pi/4$), 
whereas for $\lambda\in[\lambda_{n\pm1}^\star,\pi/2]$ only the 
second term contributes. This implies that in the system~\eqref{eq:large-alpha-d} 
the equations for $n$ odd and even decouple, and 
can be solved analytically. For instance, 
from~\eqref{eq:large-alpha-d} one obtains for the N\'eel quench 
\begin{equation}
\label{eq:large-alpha-sol}
\gamma^{\mathrm{N}}_n(\lambda) = \begin{cases}
 (s_+-s_-)\star(\ln\vartheta^\mathrm{N}_{n-1}+\ln\vartheta^\mathrm{N}_{n+1}), & n\text{ even,}\\
\begin{aligned}[b](s_+-s_-)\star\Big\{\Theta[(1-\vartheta^\mathrm{N}_{n-1})e^{\gamma^{\mathrm{N}}_{n-1}}-\vartheta^\mathrm{N}_{n-1}]\big(\ln(1-\vartheta^\mathrm{N}_{n-1})+\gamma^{\mathrm{N}}_{n-1}\big)
\\+\Theta[\vartheta^\mathrm{N}_{n-1}-(1-\vartheta^\mathrm{N}_{n-1})e^{\gamma^{\mathrm{N}}_{n-1}}]\ln \vartheta^\mathrm{N}_{n-1}
\Big\} + \left ( n\to n+2 \right ),\end{aligned} &n\text{ odd.}
\end{cases}
\end{equation}
Here $\vartheta_n^\mathrm{N}(\lambda)$ are the filling functions describing the quench from the N\'eel state. 
The Heaviside theta functions in~\eqref{eq:large-alpha-sol} fix the value of $\lambda_n^\star$. 
The two sets of equations in~\eqref{eq:large-alpha-sol} give a close-form expression 
for $\gamma_n(\lambda)$ in terms of the filling functions $\vartheta^\mathrm{N}_n(\lambda)$ describing 
the N\'eel quench. Now, while we exploited the large $\Delta$ 
expansion to derive~\eqref{eq:large-alpha-sol}, one can verify 
that~\eqref{eq:large-alpha-sol} is exact also for moderately small 
values of $\Delta$.  Precisely, upon lowering $\Delta$ the condition 
that the first Heaviside theta function in~\eqref{eq:large-alpha-d} vanishes for any $\lambda$, which ensures 
the exact solvability, remains valid, except near $\Delta=1$. 
We numerically verified that \cref{eq:large-alpha-sol} is exact 
at $\Delta\gtrsim 2.5$. 
For the dimer quench one has that $\vartheta_n\to1$ in the large $\Delta$ limit for even $n$. 
The functions $\gamma_n(\lambda)$ for the dimer quench are the same as~\eqref{eq:large-alpha-sol} 
after exchanging the even and the odd $n$. Precisely, one has 
\begin{equation}
\label{eq:large-alpha-sol-dimer}
\gamma^{\mathrm{D}}_n(\lambda)=\begin{cases}
    (s_+-s_-)\star(\ln\vartheta^\mathrm{D}_{n-1}+\ln\vartheta^\mathrm{D}_{n+1}), & n\,\text{ odd,}\\
    \begin{aligned}[b](s_+-s_-)\star\Big\{\Theta[(1-\vartheta^\mathrm{D}_{n-1})e^{\gamma^{\mathrm{D}}_{n-1}}-\vartheta^\mathrm{D}_{n-1}]\big(\ln(1-\vartheta^\mathrm{D}_{n-1})+\gamma^{\mathrm{D}}_{n-1}\big)
\\+\Theta[\vartheta^\mathrm{D}_{n-1}-(1-\vartheta^\mathrm{D}_{n-1})e^{\gamma^{\mathrm{D}}_{n-1}}]\ln \vartheta^\mathrm{D}_{n-1}
\Big\}+\left (n\to n+2 \right )\end{aligned}, &n\,\text{ even,}
\end{cases}
\end{equation}
where $\vartheta_n^\mathrm{D}(\lambda)$ are known analytically~\cite{klobas2022growth}. 
Having $\gamma_n$, we can determine $a_0$ from~\eqref{eq:sa-klobas} as
\begin{multline}
\label{eq:a0-xxz}
a_0=2\sum_n\int_+d\mu \frac{\varepsilon'_n(\mu)}{2\pi}\Big\{\ln[1-\vartheta_n(\mu)]\Theta(1-\vartheta_n(\mu)-\vartheta_n(\mu)/e^{\gamma_n(\mu)})\\
+[\ln(\vartheta_n(\mu))-\gamma_n(\mu)]\Theta(\vartheta_n(\mu)/e^{\gamma_n(\mu)}-(1-\vartheta_n(\mu)))\Big\}.
\end{multline}
In Fig.~\ref{fig:a0-xxz} we show numerical results for $a_0$ after quenches in the XXZ chain. In the figure 
we plot $a_0$ versus the chain anisotropy $\Delta$. The continuous lines are the results for the quench form the N\'eel state, whereas the dotted ones are for the dimer quench. The black curves are obtained by numerically solving the system~\eqref{eq:large-alpha-c}, and then using~\eqref{eq:a0-xxz}. The obtained $a_0$ is exact for any $\Delta$. The red curves are the explicit results~\eqref{eq:large-alpha-sol} and~\eqref{eq:large-alpha-sol-dimer} for $\gamma_n(\lambda)$ in terms of the filling functions $\vartheta_n(\lambda)$, and which are exact only for $\Delta\gtrsim 2.5$. For the N\'eel state one has that $a_0$ vanishes at large $\Delta$, reflecting that at large $\Delta$ the N\'eel state is an eigenstate of the model. On the other hand, this is not the case for the dimer quench. For the latter, we observe that $a_0$ depends mildly on $\Delta$ at large $\Delta$, similar to the von Neumann entropy~\cite{alba2018entanglement}, and to the steady-state R\'enyi entropies~\cite{mestyan2018renyi}. 

To characterize the distribution of the lower part of the entanglement spectrum, one has to determine $a_1$ (cf.~\eqref{eq:Fa}). This means that  one has to go beyond the leading order in the large $\alpha$ expansion of the R\'enyi entropies. To do that, let us consider the coupled equations~\eqref{eq:y-eq} for $y_{\alpha,n}$
\begin{equation}
\label{eq:ee1}
y_{\alpha,n}(\lambda)=\sum_{m}\int_+ d\mu [T_{nm}(\mu-\lambda)-T_{nm}(\mu+\lambda)]
\ln\left[(1-\vartheta_m(\mu))^\alpha+\frac{\vartheta^\alpha_m(\mu)}{y_{\alpha,m}(\mu)}\right]. 
\end{equation}
Let us rewrite $y_{\alpha,n}$ as 
\begin{equation}
y_{\alpha,n}(\lambda)=e^{\alpha(\gamma_n+\gamma'_n)}
\end{equation}
where we assume the $\alpha \gamma'_n\to0$ in the limit $\alpha\to\infty$. 
From \cref{eq:ee1} we obtain 
\begin{multline}
\label{eq:gamma-la}
\alpha (\gamma_n+\gamma'_n)=
\sum_{m}\int_+d\mu[T_{nm}(\mu-\lambda)-
T_{nm}(\mu+\lambda)]
\ln\left[(1-\vartheta_m(\mu))^\alpha
+\frac{\vartheta^\alpha_m}{e^{\alpha \gamma_m}}e^{-\alpha\gamma_m'}
\right]
\end{multline}
We have now to consider the large $\alpha$ limit in the integrand in~\eqref{eq:gamma-la}. To proceed, we write 
\begin{multline}
\label{eq:gamma-la-1}
\alpha (\gamma_n+\gamma'_n)=
\sum_{m}\int_+d\mu[{T}_{nm}(\mu-\lambda)-T_{nm}(\mu+\lambda)]\Big\{
\Big[
\alpha\ln(1-\vartheta_m(\mu))
\\+\ln\left(1+e^{\alpha F_{m,\infty}-\alpha\gamma'_m}\right)\Big]\Theta(\gamma'_m-F_{m,\infty})+
\Big[
\alpha\ln(\vartheta_m(\mu))-\alpha\gamma_m-\alpha\gamma'_m
+\\\ln\left(1+e^{-\alpha F_{m,\infty}+\alpha\gamma'_m}\right)\Big]\Theta(F_{m,\infty}-\gamma'_m)\Big\},
\end{multline}
where  we defined 
\begin{equation}
F_{m,\infty}(\lambda)=\ln\left(\frac{\vartheta_m}{(1-\vartheta_m)e^{\gamma_m}}\right),
\end{equation}
We can now simplify the integrals in~\eqref{eq:gamma-la} in the large $\alpha$ limit. 
At large $\alpha$ one has that $\alpha\gamma'_n$ vanishes. After expanding, we can equate the terms order by order in~\eqref{eq:gamma-la-1}. The leading order gives \cref{eq:large-alpha-c} for $\gamma_n$. At the next order we obtain a system of equations that determines $\gamma'_n(\lambda)$ as 
\begin{multline}
\label{eq:intermedio}
\gamma'_n(\lambda)=\sum_m\int_+d\mu[ {T}_{nm}(\mu-\lambda)-T_{nm}(\mu+\lambda)]\Big\{\Big[\frac{1}{\alpha}\ln(1+e^{\alpha F_{m,\infty}-\alpha\gamma_m'})\Big]\Theta(\gamma'_m-F_{m,\infty})\\
+\Big[\frac{1}{\alpha}\ln(1+e^{-\alpha F_{m,\infty}+\alpha\gamma_m'})-\gamma_m'\Big]\Theta(F_{m,\infty}-\gamma'_m)\Big\}
\end{multline}
Now, in taking the large $\alpha$ limit in the logarithms in~\eqref{eq:intermedio} we have to distinguish the case in which $F_{m,\infty}$ is vanishing at some rapidity or it is nonzero for any $\lambda$. 

If $F_{m,\infty}$ is nonzero for any $\lambda$, it is straightforward to expand the logarithms. If $F_{m,\infty}(\lambda)$ vanishes for some $\lambda^\star_m$, one has to consider the contribution of the region near  $\lambda^\star_m$ in the integrals. 
Let us summarize our findings for the quench from the N\'eel state for $\Delta>\Delta_c$.  Since we have that $F_{m,\infty}(\lambda)$ is positive for any $\lambda$ for odd $m$, whereas it has a zero for even $m$, \cref{eq:intermedio} becomes 
\begin{multline}
\gamma_n'(\lambda)=\frac{1}{\alpha^2}\frac{\pi^2}{6}\sum_{\mathrm{even}\,m}[T_{nm}(\mu_m^\star-\lambda)-T_{nm}(\mu^\star_m+\lambda)]\frac{1}{F'_{m,\infty}(\mu_m^\star)}\\
-\sum_{\mathrm{even}\,m}\int_+d\mu[T_{nm}(\mu-\lambda)-T_{nm}(\mu+\lambda)]
\gamma_m'(\mu)\Theta(-F_{m,\infty})\\
-\sum_{\mathrm{odd}\,m}\int_+d\mu[T_{nm}(\mu-\lambda)-T_{nm}(\mu+\lambda)]
\gamma_m'(\mu)\Theta(F_{m,\infty})
\end{multline}
Let us instead consider the dimer state. 
One can readily verify that  
$F_{n,\infty}(\lambda)$ is never vanishing for even $n$, whereas it vanishes for odd $n$. This is opposite to what observed for the N\'eel quench. Thus, \cref{eq:intermedio} becomes 
\begin{multline}
\gamma_n'(\lambda)=\frac{1}{\alpha^2}\frac{\pi^2}{6}\sum_{\mathrm{odd}\,m}[T_{nm}(\mu_m^\star-\lambda)-T_{nm}(\mu^\star_m+\lambda)]\frac{1}{|F'_{m,\infty}(\mu_m^\star)|}\\
-\sum_{\mathrm{odd}\,m}\int_+d\mu[T_{nm}(\mu-\lambda)-T_{nm}(\mu+\lambda)]
\gamma_m'(\mu)\Theta(-F_{m,\infty})\\
-\sum_{\mathrm{even}\,m}\int_+d\mu[T_{nm}(\mu-\lambda)-T_{nm}(\mu+\lambda)]
\gamma_m'(\mu)\Theta(F_{m,\infty}). 
\end{multline}
It is now straightforward to obtain the coefficient $a_1$. We have for the dimer quench at $\Delta>\Delta_c$ that $\gamma_n'(\lambda)=0$ for any odd $n$, which implies that 
\begin{equation}
\label{eq:a1-dimer}
a_1^{\mathrm{D}}=\frac{\pi}{3}\sum_{\mathrm{odd\,m}}\frac{\varepsilon'_m(\mu_m^\star)}{|F_{m,\infty}'(\mu_m^\star)|}-2\sum_{\mathrm{even\,m}}\int_+d\mu \frac{\varepsilon'_m(\mu)}{\pi}\gamma_m'(\mu)\Theta(F_{m,\infty}(\mu)). 
\end{equation}
For the N\'eel quench, one has instead that for $\Delta>\Delta_c$, $\gamma'_n(\lambda)=0$ for any even $n$, implying that 
\begin{equation}
\label{eq:a1-neel}
a_1^{\mathrm{N}}=\frac{\pi}{3}\sum_{\mathrm{even\,m}}\frac{\varepsilon'_m(\mu_m^\star)}{|F_{m,\infty}'(\mu_m^\star)|}-2\sum_{\mathrm{odd\,m}}\int_+d\mu \frac{\varepsilon'_m(\mu)}{\pi}\gamma_m'(\mu)\Theta(F_{m,\infty}(\mu)),  
\end{equation}
which is formally the same as for the dimer after exchanging odd and even $m$ in the sums. 
We verified that from~\eqref{eq:a1-neel} $a_1$ vanishes at any $\Delta$ for the N\'eel quench, whereas it is finite for the dimer quench. 
%
\begin{figure}
\centering
          \includegraphics[width=.8\linewidth]{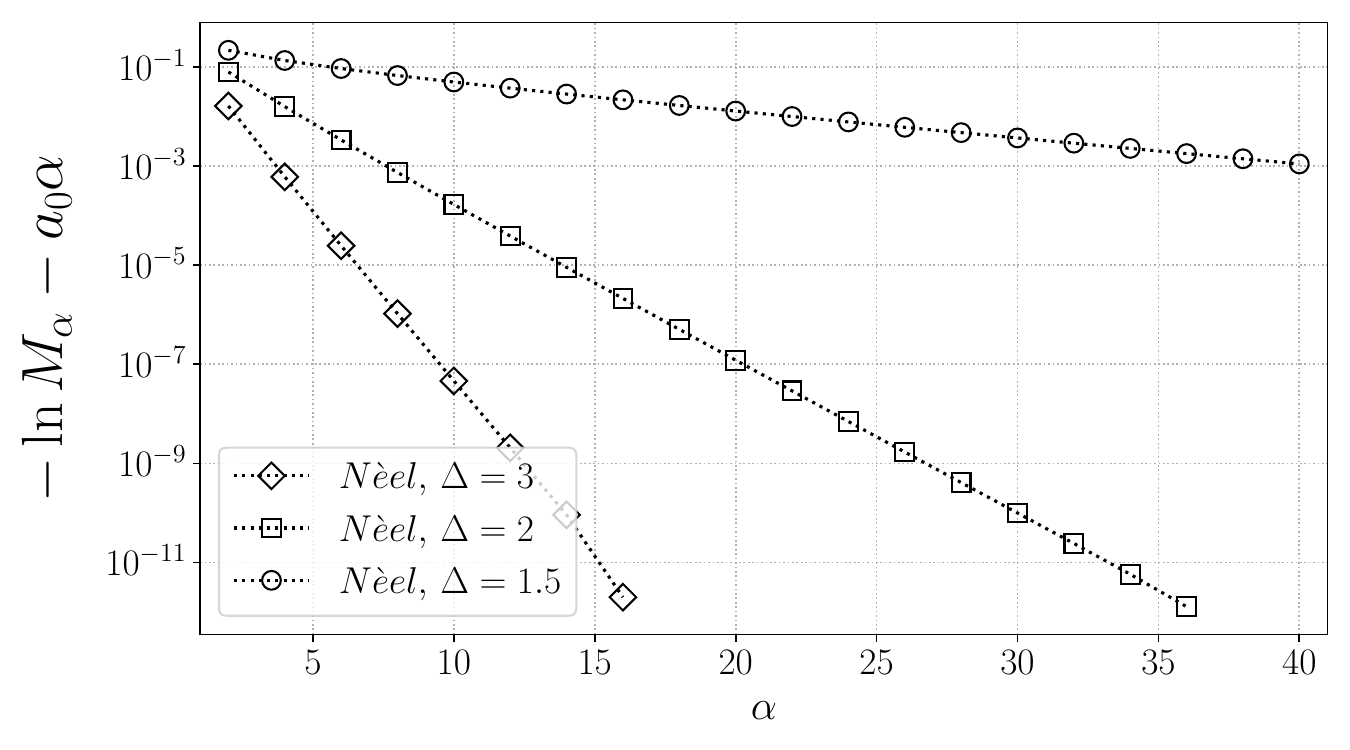}
    \caption{Large $\alpha$ behavior of $F_\alpha$ (cf.~\eqref{eq:Fa}) after the N\'eel quench in the XXZ chain. We plot $-\ln M_\alpha-a_0\alpha$, with $M_\alpha=\mathrm{Tr}\rho_A^\alpha$ and $a_0$ as obtained analytically by solving~\eqref{eq:large-alpha-c}. On the y-axis we employ a logarithmic scale. The exponential decay at large $\alpha$ is clearly visible for any $\Delta$.   
    }
    \label{fig:a1-xxz-neel}
\end{figure} 
%
To benchmark our results for $a_1$ in Fig.~\ref{fig:a1-xxz-neel}  we show numerical results for $-\ln M_\alpha-a_0\alpha$ plotted versus $\alpha$ for several values of $\Delta$. The data are obtained by numerically solving \cref{eq:y-eq} to determine $y_{\alpha,n}(\lambda)$, which are used in~\eqref{eq:sa-klobas} to obtain $M_\alpha$. The coefficient $a_0$ is obtained by numerically solving~\eqref{eq:a0-xxz}. Now, Fig.~\ref{fig:a1-xxz-neel} shows that for the N\'eel quench $-\ln M_\alpha-a_0\alpha$ decays exponentially with $\alpha$, which means that $a_1=0$, which is in agreement with expression~\eqref{eq:a1-neel}. 
%
\begin{figure}
    \centering
          \includegraphics[width=.8\linewidth]{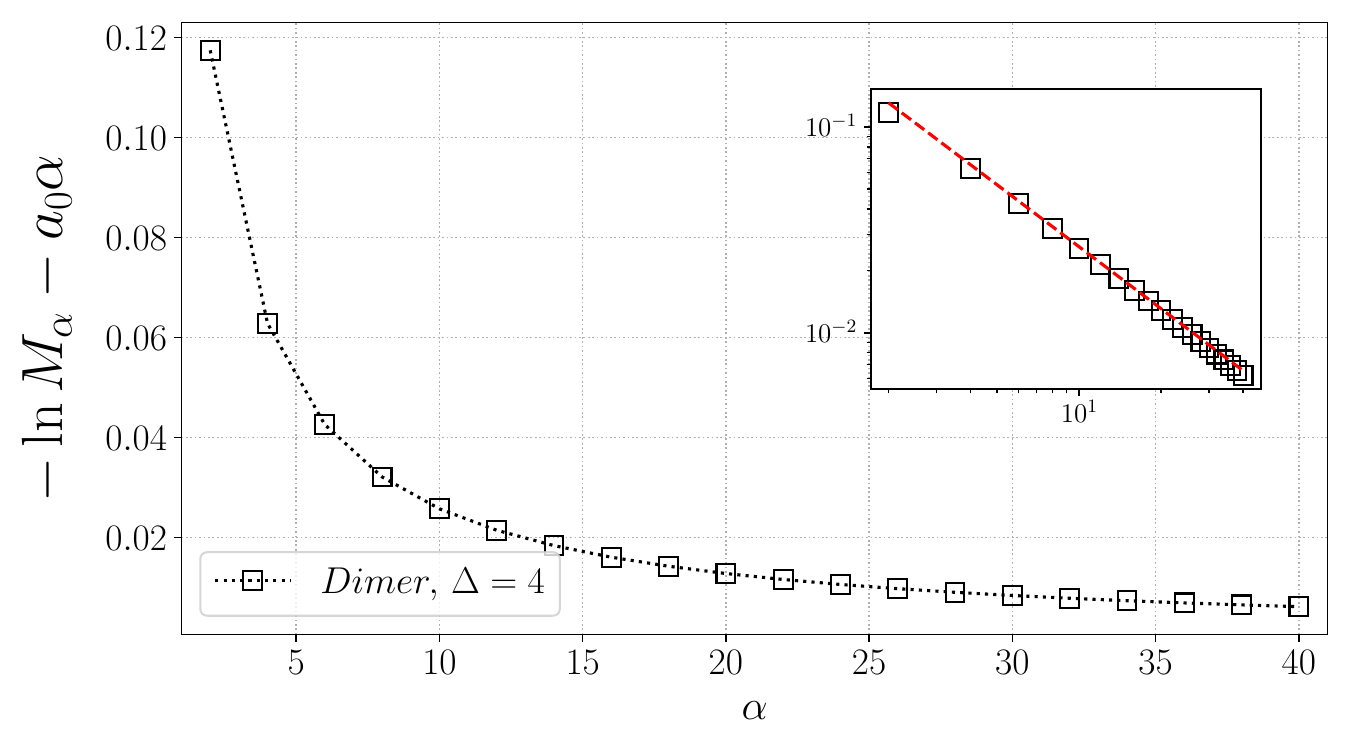}
    \caption{Dimer quench in the XXZ chain. We plot $-\ln M_\alpha-a_0\alpha$ versus $\alpha$. The data are for the dimer quench in the XXZ with $\Delta=4$. In contrast with the N\'eel quench (see Fig.~\ref{fig:a1-xxz-neel}) the large $\alpha$ behavior is power-law, as we show in the inset by using a double logarithmic scale. The dashed line is the analytical prediction for $a_1$ obtained by using~\eqref{eq:a1-dimer}. 
    }
    \label{fig:a1-xxz-mg}
\end{figure} 
%
In Fig.~\ref{fig:a1-xxz-mg} we focus on the quench from the dimer state~\eqref{eq:mg-state}. We only show data for $\Delta=4$, although we verified that the  qualitative behavior does not depend significantly on $\Delta$, at least for $\Delta>1$. In contrast with the N\'eel quench (see Fig.~\ref{fig:a1-xxz-neel}), now $-\ln M_\alpha-a_0\alpha$ decays as a power law at large $\alpha$. This is clear from the inset in Fig.~\ref{fig:a1-xxz-mg}, where we plot the same data  using a logarithmic scale on both axes. The dashed line in the inset is $a_1/\alpha$, with $a_1\approx 0.26$ as obtained from~\eqref{eq:a1-dimer}.

\subsection{The case of the XX spin chain}\label{se:XX}
\label{sec:xx-chain}

It is interesting to consider the case with $\Delta=0$ in~\eqref{eq:xxz-ham}, for which both the scenarios outlined in Section~\ref{sec:theory} occur.  
For $\Delta=0$ \cref{eq:xxz-ham} becomes the XX spin chain, which is described by the Hamiltonian  
\begin{equation}
\label{eq:xx-ham}
    H=-\frac{1}{2} \sum_{j = 1}^{N} ( \px_j \px_{l+j}+\py_j\py_{j+1}). 
\end{equation}
The Hamiltonian~\eqref{eq:xx-ham} is mapped to a free-fermion Hamiltonian via the Jordan-Wigner transformation as 
\begin{equation}
\label{eq:ham-xx-1}
    H=-\sum_{j} (a_{j}^{\dagger} a_{j+1}+a_{j+1}^{\dagger} a_{j}),
\end{equation}
where $a_j$ are standard Dirac fermion operators. 
By a discrete Fourier transformation \cref{eq:ham-xx-1} is diagonalized as 
\begin{equation}
 H=-\sum_{k} \varepsilon_k  c_{k}^{\dagger} c_{k},   
\end{equation}
where $c_k$ are the Fourier transform of $a_j$, and satisfy standard fermionic anticommutation relations. 
The dispersion relation is given as 
\begin{equation}
 \varepsilon_{k}=\cos (k). 
\end{equation}
As for the TFIC, by using Peschel's trick~\cite{peschel2009reduced} the entanglement spectrum is obtained from the correlation matrix $C_{ij}$ defined as 
\begin{equation}
    C_{ij}=\langle c_{i}^{\dagger}(t)c_{j}(t)\rangle
\end{equation}
For both the N\'eel quench and the dimer quench the matrix $C_{ij}$ is known analytically (see, for instance, Ref.~\cite{parez2025reduced}). 
The fermionic N\'eel state is defined as 
\begin{equation}
|N\rangle=|01010\cdots\rangle. 
\end{equation}
The time-dependent correlation matrix  is given as   
\begin{equation}
    \langle N|c_i^\dagger(t)c_j(t)|N\rangle=\frac{1}{2}\delta_{i,j}+\frac{(-1)^j}{2}\int_{-\pi}^\pi \frac{dk}{2\pi}e^{ \mathrm{i}k(i-j)+4\mathrm{i}t\cos(k)},
\end{equation}
where we considered the thermodynamic limit $L\to\infty$. 
The fermionic dimer state is defined as 
\begin{equation}
|D\rangle=\prod_{j=1}^{L/2} \frac{a_{2j}^\dagger-a_{2j-1}^\dagger}{\sqrt{2}}|0\rangle. 
\end{equation}
It is convenient to exploit the two-site translation invariance rewriting the correlation matrix as 
\begin{equation}\label{eq:JAdim}
    J(t) =2C-\mathbf{1}= \begin{pmatrix}
\pi_0(t)& \pi_1(t) & \cdots & \pi_{\ell/2-1}(t)\\
\pi_{-1} (t)& \pi_0(t) &  &  \vdots\\
\vdots & &\ddots &\vdots \\
\pi_{1-\ell/2} (t)&\cdots &\cdots &\pi_0(t)
\end{pmatrix}.
\end{equation}
%
%
\begin{figure}
    \centering
          \includegraphics[width=.8\linewidth]{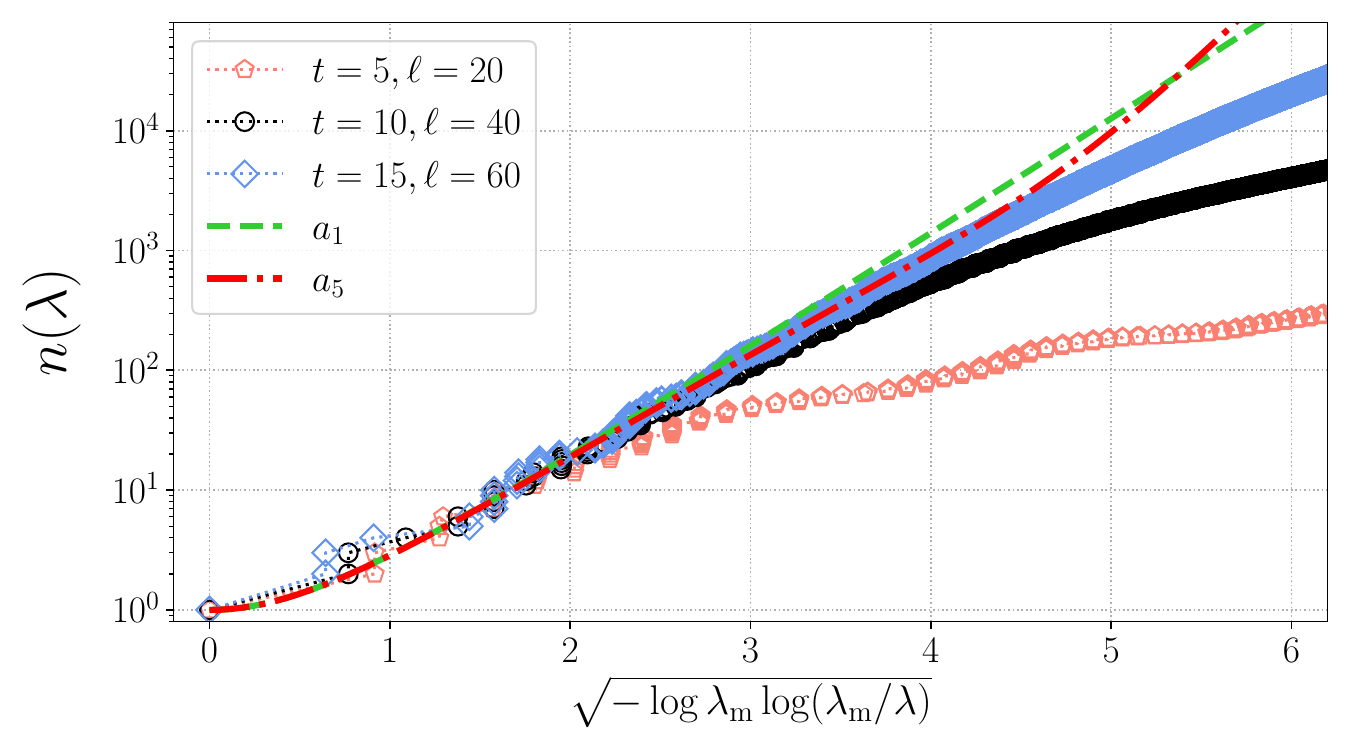}
    \caption{
    Cumulative distribution $n(\lambda)$ of the entanglement spectrum levels after the magnetic field quench $h_0\to h$ in the TFIC. Here we choose $h=1$ and  $h_0=0$. The symbols are exact numerical data for a subsystem of size $\ell$ embedded in an infinite chain. The different symbols are for different times $t = 5, 10$ and $15$ and $\ell$, with the ratio $t/\ell=1/4$ fixed. The dashed and dashed-dotted lines are the theoretical predictions in the large $t,\ell$ limit (cf.~\eqref{eq:n-fin-fin}) obtained by including the  terms up to $\mathcal{O}(b^0)$ and $\mathcal{O}(b^{-4})$,
    respectively. 
    }
    \label{fig:ising-2}
\end{figure} 
%
In the thermodynamic limit $L\to \infty$, the blocks $\pi_j(t)$ 
read~\cite{parez2025reduced}
\begin{equation}
    \pi_{m}(t) =  \ \begin{pmatrix}
-f_m(t)&-g_m(t)\\
 -g_{-m}(t)&f_m(t) \\
\end{pmatrix}
\end{equation}
with
\begin{equation}
\left\{
\begin{aligned}
f_m(t) &=  \int_{-\pi}^{\pi}\frac{d k}{2 \pi}e^{-2i m k} \sin (k) \sin(2 \cos(k) t), \\
g_m(t) &=  \int_{-\pi}^{\pi}\frac{d k}{2 \pi}e^{-2i m k} e^{-i k}(\cos (k) + i \sin (k) \cos(2 \cos(k) t)).
\end{aligned}
\right.
\end{equation}
Now, the R\'enyi entropies are obtained from the correlator $C_{ij}$ restricted to subsystem $A$~\cite{peschel2009reduced}. The dynamics of the R\'enyi entropies is described within 
the quasiparticle picture. For the N\'eel and the dimer quench it is straightforward to show that the moments of the reduced density matrix are of the same form as~\eqref{eq:ln-tr}. Precisely, one obtains  
\begin{equation}
\ln(\mathrm{Tr}\rho_A^\alpha)=\int_{-\pi}^\pi\frac{dk}{2\pi}
\left\{\alpha\ln\left[\frac{1+D_k}{2}\right]+\ln\left[1+\left(\frac{1-D_k}{1+D_k}\right)^\alpha\right]\right\}\min(|v_k|t,\ell). 
\end{equation}
For the N\'eel quench one has that $D_k=0$, whereas for the dimer quench 
it is $D_k=\cos(k)$. This implies that for the N\'eel quench $M_\alpha=2^{-\alpha}$, i.e., $a_j=0$ for any $j\ge 1$ in \cref{eq:Fa}, which means that $P(\lambda)\sim \delta(\lambda-\lambda_\mathrm{m})$ (see Section~\ref{sec:staircase}). On the other hand, 
for the dimer quench one has that $a_1\ne 0$, implying that in the long time limit at fixed $\xi$ the lower part of the entanglement spectrum is described by~\eqref{eq:P-intro}. We do not report the explicit formula for $a_1$, which can be obtained by employing the same strategy as in Section~\ref{sec:ising}. 

\section{Numerical benchmarks}
\label{sec:numerics}

%
\begin{figure}
    \centering
          \includegraphics[width=.8\linewidth]{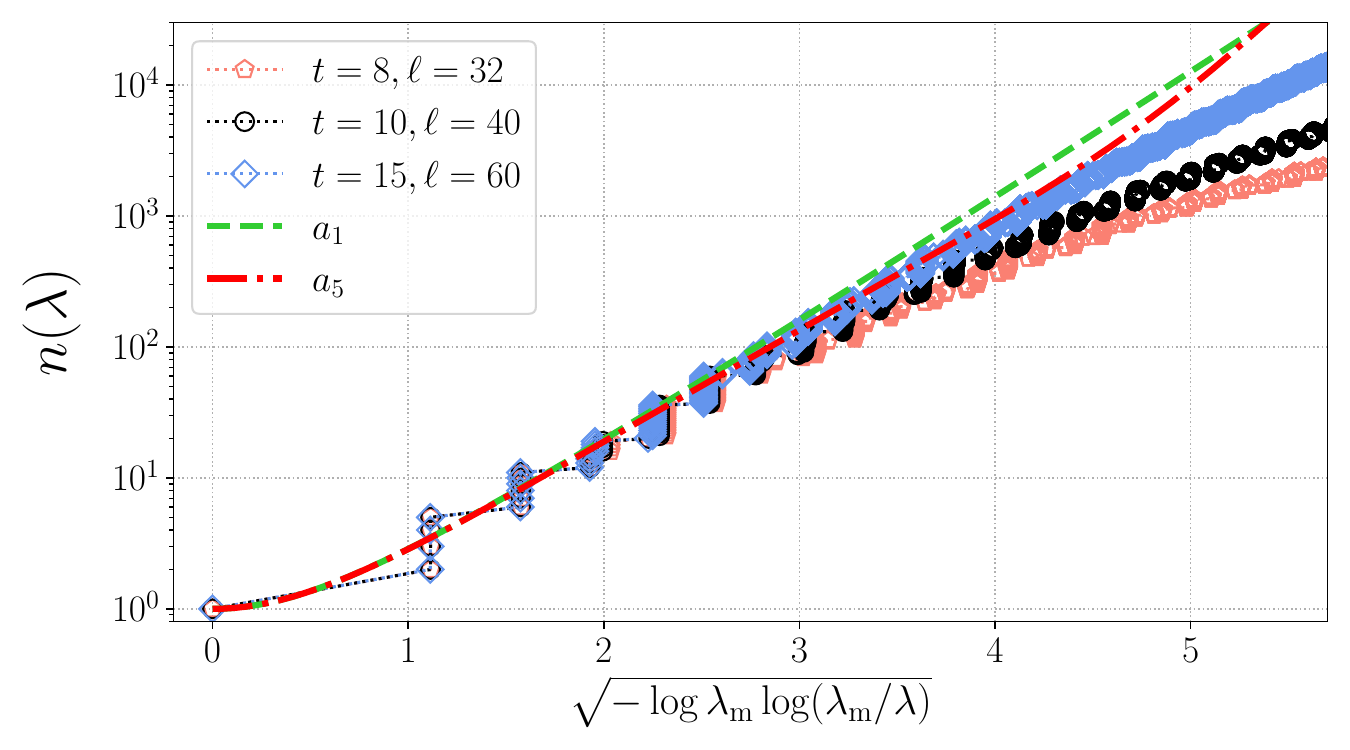}
    \caption{Cumulative distribution $n(\lambda)$ of the eigenvalues of the reduced density matrix after the quench from the dimer state in the XX chain. We plot $n(\lambda)$ versus the scaling variable $\xi=[-\ln\lambda_\mathrm{m}\ln(\lambda_\mathrm{m}/\lambda)]^{1/2}$, where $\lambda_\mathrm{m}$ is the largest eigenvalue. Notice the logarithmic scale in the $y$-axis. The different symbols are exact numerical data for different subsystem sizes $\ell$ and times $t$. We consider data at fixed ratio $t/\ell=1/4$. The dashed line is the theoretical prediction~\eqref{eq:n-fin-fin} including only the contribution of $a_1$. The dashed-dotted line is~\eqref{eq:n-fin-fin} where we include the contributions of $a_3, a_5$. 
    }
    \label{fig:xx-dimer}
\end{figure} 
%
Here we discuss numerical data supporting the scenarios described in the previous sections. Precisely, in Section~\ref{sec:cft-like} we focus on the entanglement spectrum after quenches that give rise to CFT-like behaviors. We consider both free-fermion models, such as the TFIC and the XX chain, as well as the XXZ spin chain, which is an interacting model. In Section~\ref{sec:staircase} we discuss quenches giving rise to the staircase scenario described in Section~\ref{sec:a1-van}. 

\subsection{CFT-like scenario}
\label{sec:cft-like}

We start considering quenches that give rise to a CFT-like structure in the entanglement spectrum. Precisely, we start considering quenches to the critical point in the TFIC (see Section~\ref{sec:ising}). We focus on the short-time regime. In Fig.~\ref{fig:ising-2} we plot the cumulative distribution function $n(\lambda)$ (cf.~\eqref{eq:nlambda}) versus the scaling variable $\xi$ (cf.~\eqref{eq:P-intro}). We consider the quenches $h_0\to h$ with $h=1$ (critical point) and $h_0=0$. The symbols in the figure are results for different subsystem lengths $\ell=20,40,60$ and different times. We consider fixed ratio $t/\ell=1/4$ to ensure that we are in the short-time regime.  We checked that the scenario remains the same for other values of $t/\ell$ in the short-time regime. The data exhibit collapse upon increasing time. 
The dashed line in the figure is~\eqref{eq:n-intro}, which holds at $t\to\infty$ for any fixed $\xi$. In the figure we also report with the dashed-dotted line \cref{eq:n-fin-fin}, in which we include the contribution of the term $a_3/\alpha^3$ and $a_5/\alpha^5$ in the expansion of $F_\alpha$ (cf.~\eqref{eq:Fa}). At small values of $\xi$ the two theoretical predictions are indistinguishable. At larger $\xi$, differences are visible, although they vanish at long times. The agreement between the numerical data and the theoretical curve is excellent up to $\xi\approx 5$. 
%
\begin{figure}
    \centering
          \includegraphics[width=.8\linewidth]{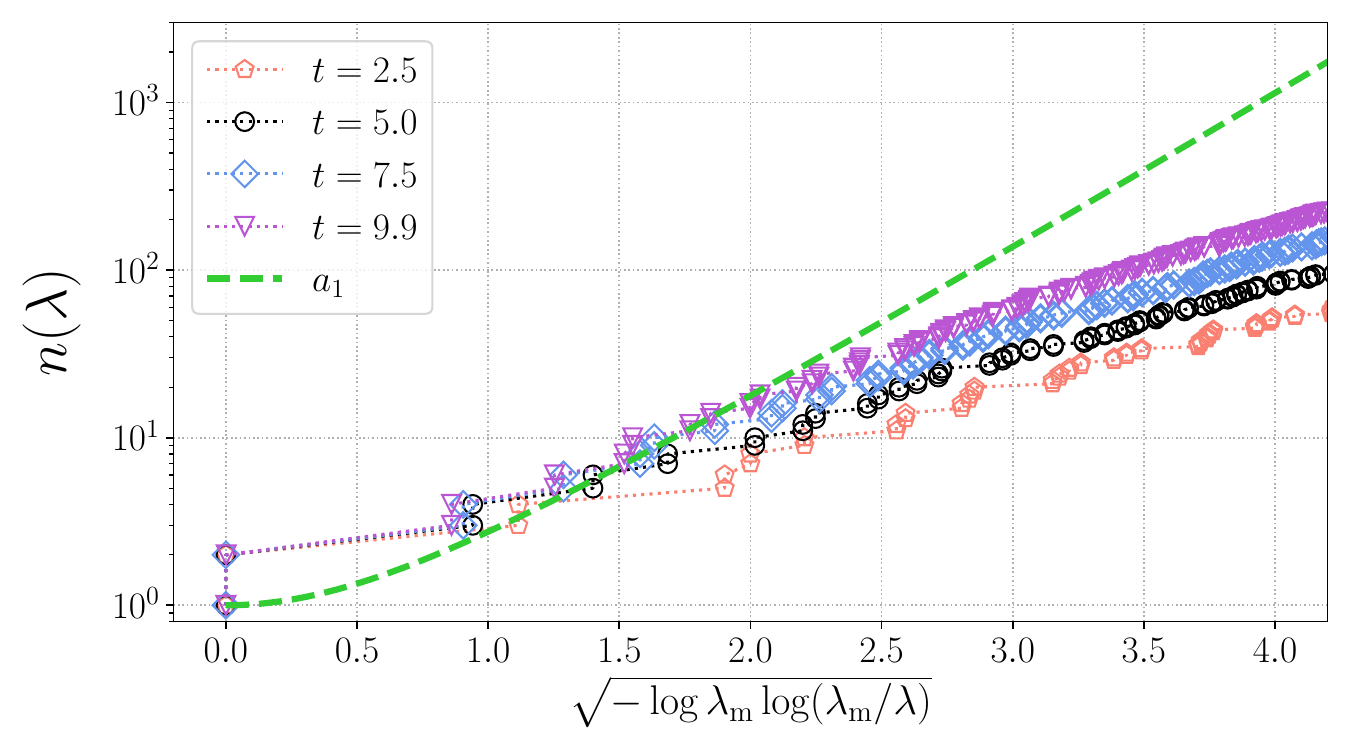}
    \caption{Cumulative distribution $n(\lambda)$ of the eigenvalues of the reduced density matrix after the quench from the dimer state in the XXZ chain at $\Delta=4$. We plot TEBD data for $n(\lambda)$ versus the scaling variable $\xi=[-\ln\lambda_\mathrm{m}\ln(\lambda_\mathrm{m}/\lambda)]^{1/2}$, where $\lambda_\mathrm{m}$ is the largest eigenvalue. Notice the logarithmic scale in the $y$-axis. The different symbols are exact numerical data for different subsystem sizes $\ell$ and times $t$. The results are in the short-time regime. The dashed line is~\eqref{eq:n-fin-fin} including only the term with $a_1$.  
    }
    \label{fig:xxz-dimer}
\end{figure} 
%
As a second example of out-of-equilibrium entanglement spectrum that exhibits the CFT-like structure we consider in Fig.~\ref{fig:xx-dimer} the quench from the dimer state~\eqref{eq:mg-state} in the XX chain. In the figure we show numerical data for $n(\lambda)$ as a function of $\xi$. We show entanglement spectra at times $t=8,10,15$ and fixed ratio $t/\ell=1/4$. Similar to the case of the TFIC, upon increasing time the data exhibit collapse on the same curve. The dashed and dashed-dotted lines are the theoretical predictions in the limit $t\to\infty$ at fixed $\xi$, which are obtained by including in~\eqref{eq:n-fin-fin} the terms $\simeq 1/b^2$ and $1/b^4$. At small values of $\xi$ the dashed line captures quite well the behavior of the numerical data. At moderately large values of $\xi\simeq 4$ the dashed-dotted line is closer to the data, i.e., it captures also the subleading corrections. 

In Fig.~\ref{fig:xxz-dimer} we consider the interacting case, focusing on the quench from the dimer state in the XXZ chain at $\Delta=4$. 
Fig.~\ref{fig:xxz-dimer} shows time-dependent Density Matrix Renormalization Group (tDMRG) data~\cite{schollwoeck2011the,paeckel2019time} for the entanglement spectrum (see Ref.~\cite{klobas2022growth} for the details on the simulation). The tDMRG simulations are performed by writing the initial state in the Matrix Product State (MPS) form. Both the N\'eel and the dimer states  admit a MPS representation with small bond dimension $\chi$. Then we apply a second-order  Trotter decomposition of the evolution operator. We verified that a time-step $\delta t=0.1$ is sufficient to obtain accurate results.  During the dynamics the bond dimension $\chi$ of the MPS increases exponentially with time. For this reason, at each time step we truncate the MPS by performing a Singular Value Decomposition (SVD) keeping the largest $\chi_\mathrm{max}=2048$ singular values. This allows us to reach times $t\lesssim 10$.  In contrast with the dimer quench in the XX chain, the lowest level of the spectrum, i.e., the largest eigenvalue of the reduced density matrix is doubly degenerate. Moreover, the finite-time corrections are ``large'' at $t=9.9$, which is the largest time that we can simulate, and the data exhibit a sizable drift, even at small values of $\xi$. The dashed line is the theoretical prediction, which is obtained from~\eqref{eq:n-fin-fin}, where we keep only the leading contribution, i.e., the first term in~\eqref{eq:n-fin-fin}. The parameters $a_0,a_1$ are obtained from~\eqref{eq:a0-xxz} and~\eqref{eq:a1-dimer}. For $\Delta=4$ we have 
$a_0\approx 0.2$ and $a_1\approx 0.26$. In contrast with the XX case (see Fig.~\ref{fig:xx-dimer}) the agreement  between the numerics and~\eqref{eq:n-fin-fin} is not perfect, although the discrepancy can be attributed to finite-time effects.

\subsection{The staircase scenario}
\label{sec:staircase} 

\begin{figure}
    \centering
          \includegraphics[width=.45\linewidth]{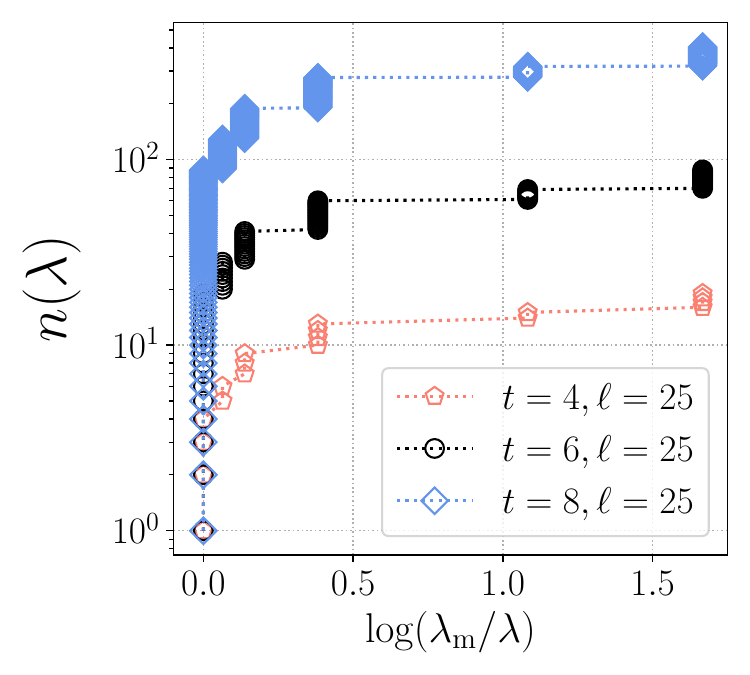}
        \includegraphics[width=.45\linewidth]{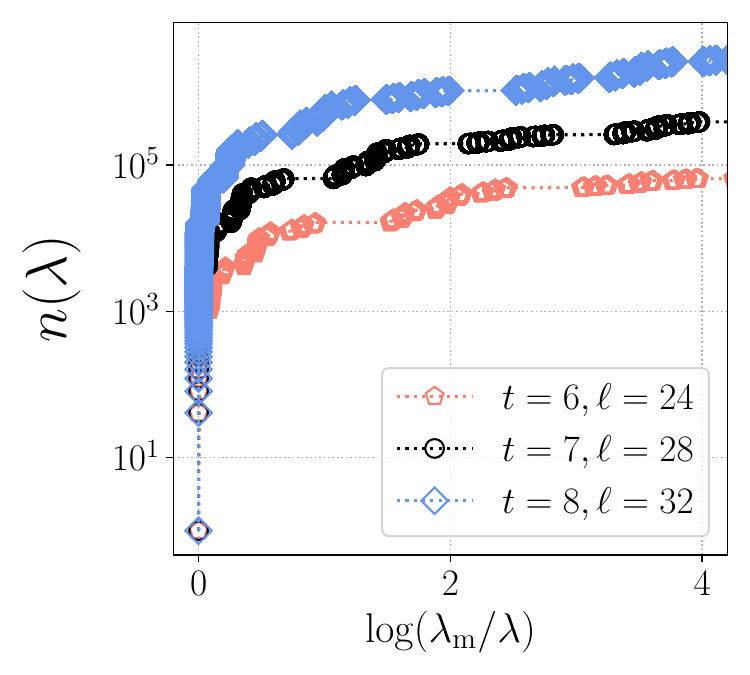}
    \caption{(Left panel) Cumulative distribution $n(\lambda)$ after the quench from the state $|\Psi_0\rangle$ with $\vartheta=\vartheta_c$ (cf.~\eqref{eq:psi0-54}) in the rule $54$ chain. 
    We plot $n(\lambda)$ versus $\ln(\lambda_\mathrm{m}/\lambda)$, with $\lambda_\mathrm{m}$ the largest eigenvalue of $\rho_A$. The different symbols are for different times. Notice the step-like behavior at $\lambda=\lambda_\mathrm{m}$. (Right panel) Same as in the left panel for the N\'eel quench in the XX chain. In both cases the step-like form is attributed to pure linear behavior in $\alpha$ of the moments $M_\alpha=a_0\alpha$. 
    }
    \label{fig:rule54}
\end{figure}

Let us now discuss the situation in which $F_\alpha=a_0\alpha$ (cf.~\eqref{eq:Fa})  apart from exponentially suppressed terms in the large $\alpha$ limit. This gives rise to the staircase pattern in $n(\lambda)$ that was discussed in Section~\ref{sec:a1-van}.  The scenario occurs for quenches in the TFIC with $h\ne 1$, for the N\'eel quench in the XX and the XXZ chain, and for quenches in the rule $54$ chain (see Section~\ref{sec:rule-54}). 
%
We start discussing the quench from the state~\eqref{eq:psi0-54} in the rule $54$ chain. Our results are summarized in Fig.~\ref{fig:rule54}. For any value of the quench parameter $0< \vartheta<1$, one has that $F_\alpha=a_0\alpha+a_1e^{-d_1\alpha}+\dots$ (see Section~\ref{sec:rule-54}), where the dots stand for exponentially suppressed terms in the large $\alpha$ limit. Moreover, for generic $\vartheta$ one has $a_1\ne0$, whereas $a_1$ vanishes at $\vartheta=\vartheta_c\approx 0.31$. Let us first consider the situation with $a_1=0$, i.e., quenches from~\eqref{eq:psi0-54} with $\vartheta=\vartheta_c$. According to the results of Section~\ref{sec:rule-54}, since $a_1$ vanishes, the probability distribution function $P(\lambda)$ of the eigenvalues of the reduced density matrix becomes a delta function, i.e., 
\begin{equation}
\label{eq:P-delta}
P(\lambda)=\lambda_\mathrm{m}^{-1}\delta(\lambda-\lambda_\mathrm{m}), 
\end{equation}
which implies that the cumulative distribution is 
%
$n(\lambda)=\lambda_\mathrm{m}^{-1}\Theta(\lambda_\mathrm{m}-\lambda)$.~\cref{eq:P-delta} implies that most of the eigenvalues of the reduced density matrix become degenerate in the long-time limit. 
This is supported in the left panel of  Fig.~\ref{fig:rule54}  where we show tDMRG data for the quench in the rule $54$ from the state~\eqref{eq:psi0-54} with $\vartheta=\vartheta_c$. We show  data for $\ell=25$ and $t=4,6,8$, which are reported with different symbols. In the figure we plot $n(\lambda)$ as a function of $\ln(\lambda_\mathrm{m}/\lambda)$. We checked that for all the times considered we are in the regime of linear entanglement growth, i.e., in the short-time regime. As it is clear from the figure, $n(\lambda)$  exhibits the step-like structure $\Theta(\lambda_\mathrm{m}-\lambda)$. As time increases, the number of nonzero eigenvalues of the reduced density matrix increases, although they are degenerate with $\lambda_\mathrm{m}$. Notice that there are extra eigenvalues at $\ln(\lambda_\mathrm{m}/\lambda)>0$. They signal the presence of finite-time corrections that are beyond the hydrodynamic regime. Indeed, to derive the distribution of the entanglement spectrum levels we employed the scaling~\eqref{eq:sa-klobas}, which holds in the hydrodynamic limit. In Fig.~\ref{fig:rule54} (right panel) we show results for the N\'eel quench in the XX chain. The qualitative behavior is the same as for the rule $54$ chain. The largest eigenvalue $\lambda_\mathrm{m}$ becomes highly degenerate and $n(\lambda)$ exhibits a step-like behavior. We observe that the number of eigenvalues $\lambda<\lambda_\mathrm{m}$, i.e., that violate~\eqref{eq:P-delta} is larger for the XX chain. 
  
\begin{figure}
    \centering
        \includegraphics[width=.45\linewidth]{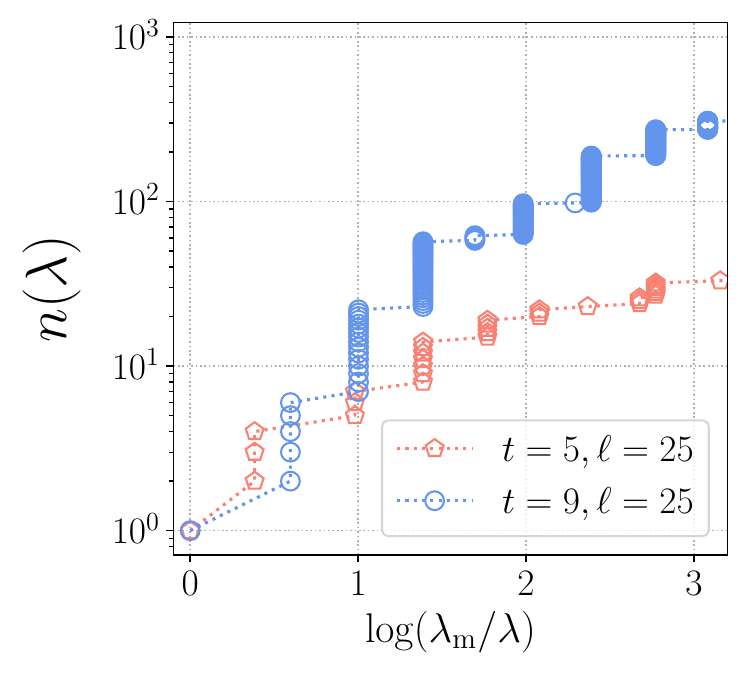}
         \includegraphics[width=.45\linewidth]{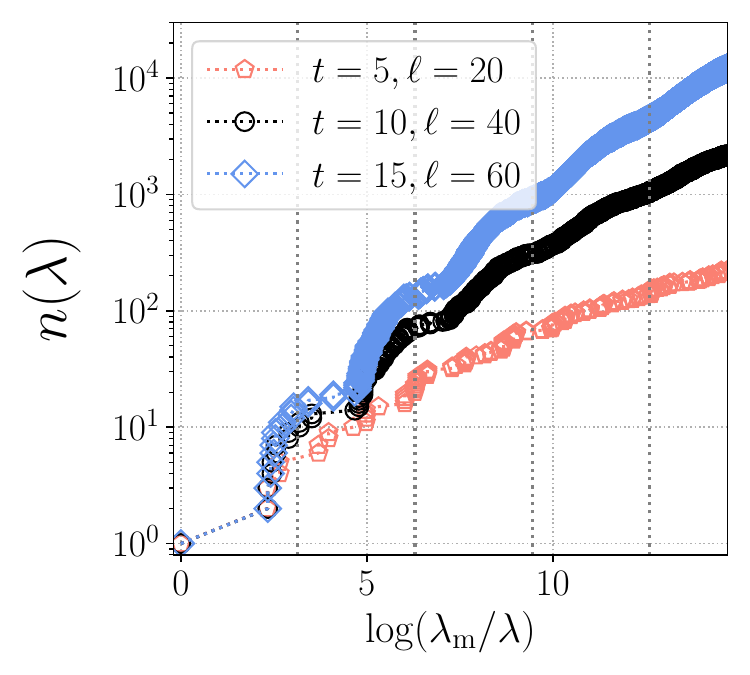}
    \caption{Cumulative distribution $n(\lambda)$ versus $\ln(\lambda_\mathrm{m}/\lambda)$  after quenches in the rule $54$ chain (left panel) and in the TFIC (right panel). For the rule $54$ chain we consider the quench from the initial state $|\Psi_0\rangle$ with $\vartheta=1/2$. For the TFIC we focus on the quench $h_0\to h$, with $h_0=8$ and 
    $h=2$. The different symbols in the figures are data for different times. In both quenches the behavior of the moments $M_\alpha$ is $M_\alpha=a_0\alpha +\mathcal{O}(e^{-d_1\alpha})$. This is reflected in the ``staircase'' structure of $n(\lambda)$. 
    }
    \label{fig:rule-54-ising}
\end{figure}

\begin{figure}
    \centering
          \includegraphics[width=.8\linewidth]{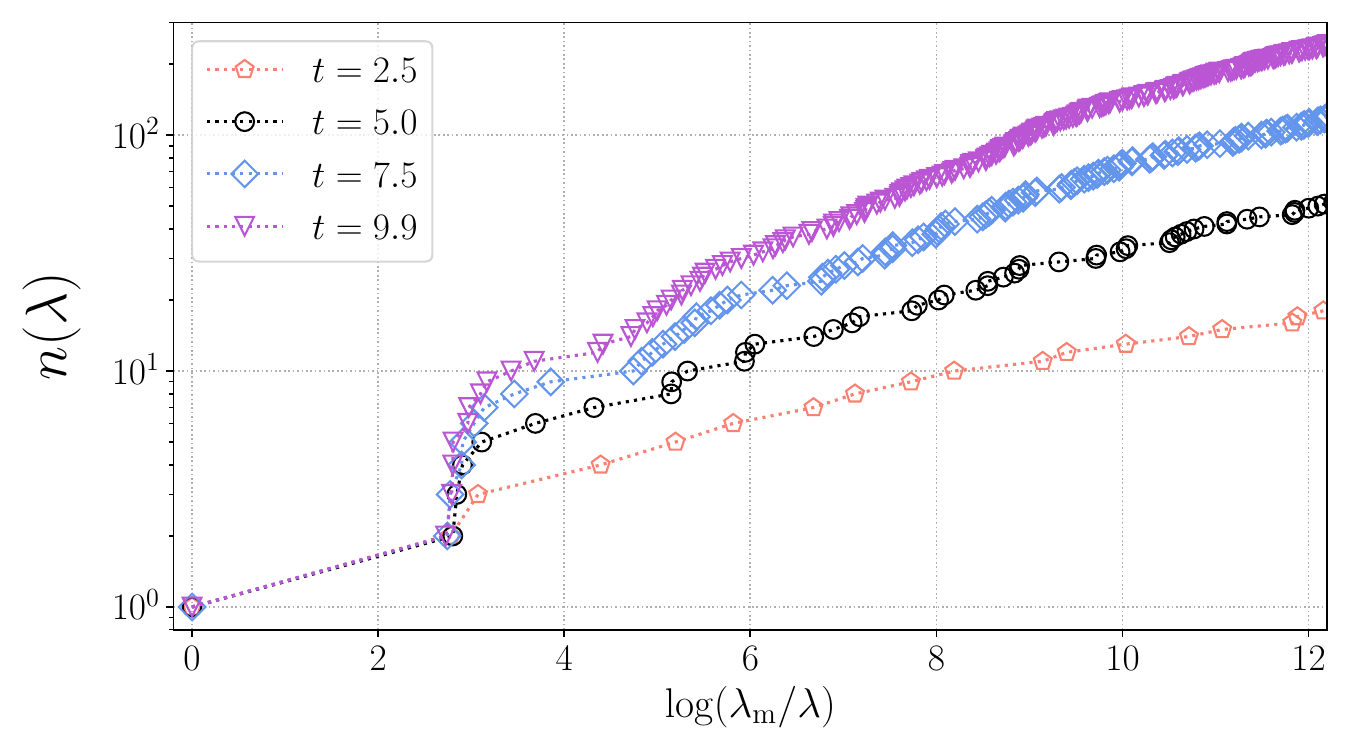}
    \caption{Cumulative  distribution $n(\lambda)$ after the quench from the 
    N\'eel state in the XXZ chain at $\Delta=4$. The symbols are tDMRG data for different times up to $t\lesssim 10$. We plot $n(\lambda)$ versus $\ln(\lambda_\mathrm{m}/\lambda)$ with $\lambda_\mathrm{m}$ the largest eigenvalue of the reduced density matrix. The data exhibit the expected staircase structure discussed in Section~\ref{sec:a1-van}.  
    }
    \label{fig:xxz-neel}
\end{figure}

Let us now discuss the scenario in which $F_\alpha$ decays exponentially in the limit $\alpha\to\infty$. Thus, let us assume that at large $\alpha$ we have  
\begin{equation}
F_\alpha=a_0\alpha+a_1e^{-d_1 \alpha}.
\end{equation}
Now, one should expect that 
\begin{equation}
n(\lambda)=\sum_{k=0}^\infty (-b)^k/k! (a_1/a_0)^k e^{d_1k}\delta(e^{-d_1 k}\lambda_\mathrm{m}-\lambda)
\end{equation}
(cf.~\eqref{eq:nl-exp}), with $b=-\ln(\lambda_\mathrm{m})$. This means that when plotted as a function of $\ln(\lambda_\mathrm{m}/\lambda)$, $n(\lambda)$ exhibits a staircase structure, with equally-spaced steps. In Fig.~\ref{fig:rule-54-ising} (left) we consider quenches in the rule $54$ chain from initial states with $\vartheta\ne \vartheta_c$. Precisely, we consider $\vartheta=1/2$. 
The circles and the pentagons in the figure are data for $t=5$ and $t=9$.
While the result exhibits a staircase-like behavior similar to the predicted one, we see that the position of the ``steps'' is different for the two times.
This is in contrast with the prediction \cref{eq:nl-exp}, according to which the first step is expected at $\ln(\lambda_\mathrm{m}/\lambda)\approx 0.46$ for all times.
Moreover, we see that the steps are not equally spaces, differently from what we expect from the prediction.
The discrepancy could be attributed to finite-time corrections. Moreover, another explanation could be that to correctly describes the structure of the entanglement spectrum one has to go beyond the hydrodynamic regime. 
%
%
In Fig.~\ref{fig:rule-54-ising} (right)  we consider the quench in the TFIC with $h_0=8$ and $h=2$. In contrast with the quenches discussed in Section~\ref{sec:cft-like}, $F_\alpha$ decays exponentially at large $\alpha$. This implies that $n(\lambda)$ exhibits the staircase structure, as confirmed in Fig.~\ref{fig:rule-54-ising}. Although the scenario is qualitatively the same as in the rule $54$ chain, only the first two steps of the staircase structure are visible. Moreover, the steps are smeared. This is likely due to the fact that the TFIC exhibits a richer dispersion as compared with the rule $54$ chain, which contains only left and right movers with unit velocity. Finally, let us consider the interacting case. In Fig.~\ref{fig:xxz-neel} we show tDMRG data for $n(\lambda)$ after the quench from the N\'eel state. As we discussed in Section~\ref{sec:theory}  one has for large $\alpha$, $F_\alpha=a_0\alpha+a_1/\alpha$, with $a_1=0$. Moreover, we checked that at large $\alpha$, $F_\alpha$ vanishes exponentially. This is consistent with the step-like structure of $n(\lambda)$, which is visible in Fig.~\ref{fig:xxz-neel}.

\section{Conclusions}
\label{sec:concl}
%
%

We investigated the distribution of the eigenvalues of the reduced density matrix (entanglement spectrum) after a global quantum quench in several one-dimensional quantum many-body systems. Several scenarios emerge, depending on the large $\alpha$ behavior of the R\'enyi entropies, equivalently the moments of the reduced density matrix. Precisely, we showed that if $S_\alpha$ contains power-law decaying terms at large $\alpha$ the distribution of the entanglement spectrum exhibits scaling behavior when plotted as a function of $\xi=\sqrt{-b\ln(\lambda_\mathrm{m}/\lambda)}$, with $b=-\ln(\lambda_\mathrm{m})$, and $\lambda_\mathrm{m}$ the largest eigenvalue of the reduced density matrix. Remarkably, at fixed $\xi$ in the long time limit the distribution of the entanglement spectrum depends only on two parameters, which are obtained from the leading and next-to-leading behavior of $S_\alpha$ in the large $\alpha$ limit. Oppositely, if $S_\alpha$ decay exponentially  the cumulative distribution $n(\lambda)$ of the entanglement spectrum exhibits a multi-step structure. We checked our results in  several models, such as the rule $54$ chain, the transverse field Ising chain, and the $XXZ$ chain. 

Let us discuss possible directions for future work. First, the scenarios that we outlined in the paper rely only on the scaling of the R\'enyi entropies. This suggests that they could be generalized to quantum quenches in higher dimensional models. It would be interesting to investigate the validity of our results beyond one dimension. While this is challenging for interacting systems, it should be feasible for models that are mappable to free fermions or free bosons. Furthermore,  our results could apply to the distribution of the ground-state entanglement spectrum levels in higher-dimensional systems. It would be also interesting to investigate whether and in which regime the scaling behavior that we derive holds for quantum quenches in random unitary circuits.  It would be also useful to understand how our findings are reflected in the structure of the entanglement Hamiltonian, for instance by exploiting the results of Refs.~\cite{rottoli2025entanglement,Travaglino:2024sxj,Travaglino:2025gix}. An intriguing direction is to further investigate the structure of the entanglement spectrum and of the entanglement Hamiltonian for quenches with $a_1=-a_0$, i.e., for which the entanglement spectrum possesses the same structure as in CFT. Finally, it would be interesting to understand if the CFT-like structure in the entanglement spectrum is related to the CFT-like for of the tripartite information uncovered in Ref.~\cite{maric2023universality,Maric2024entanglement}

\section*{Acknowledgements}

This study was carried out within the National Centre on HPC, Big Data and Quantum Computing - SPOKE 10 (Quantum Computing) and received funding from the European Union Next- GenerationEU - National Recovery and Re- silience Plan (NRRP) – MISSION 4 COMPONENT 2, INVESTMENT N. 1.4 – CUP N. I53C22000690001. This work has been supported by the project ``Artificially devised many-body quantum dynamics in low dimensions - ManyQLowD'' funded by the MIUR Progetti di Ricerca di Rilevante Interesse Nazionale (PRIN) Bando 2022 - grant 2022R35ZBF. 
G.L.  acknowledges the support by P1-0044 program of the Slovenian Research Agency, the QuantERA grant QuSiED by MVZI, QuantERA II JTC 2021, and ERC StG 2022 project DrumS, Grant Agreement 101077265.

\bibliography{bibliography.bib}

\end{document}